\documentclass[fleqn,usenatbib]{mnras}

\usepackage{newtxtext,newtxmath, comment}

\usepackage[T1]{fontenc}

\DeclareRobustCommand{\VAN}[3]{#2}
\let\VANthebibliography\thebibliography
\def\thebibliography{\DeclareRobustCommand{\VAN}[3]{##3}\VANthebibliography}

\usepackage{graphicx}	
\usepackage{amsmath}	

\title[Characterising Cloudy Mini-Neptunes with JWST]{Characterising Atmospheres of Cloudy Temperate Mini-Neptunes with JWST}

\author[S. Constantinou et al.]{
Savvas Constantinou$^{1}$\thanks{E-mail: sc938@ast.cam.ac.uk},
Nikku Madhusudhan$^{1}$\thanks{E-mail: nmadhu@ast.cam.ac.uk}
\\
$^{1}$Institute of Astronomy, University of Cambridge, Madingley Road, Cambridge CB3 0HA, UK\\
}

\date{Accepted 2022 April 27. Received 2022 April 25; in original form 2021 October 11}

\pubyear{2022}

\begin{document}
\label{firstpage}
\pagerange{\pageref{firstpage}--\pageref{lastpage}}
\maketitle

\begin{abstract}
The upcoming James Webb Space Telescope (JWST) promises a generational shift in the study of temperate mini-Neptune atmospheres using transit spectroscopy. High-altitude clouds however threaten to impede their atmospheric characterisation by muting spectral features. In this study, we systematically investigate JWST instrument configurations for characterising cloudy mini-Neptune atmospheres, assessing the importance of instrument choice and wavelength coverage, focusing on  NIRISS and NIRSpec. We consider two temperate mini-Neptunes orbiting nearby M dwarfs, K2-18~b and TOI-732~c,  with equilibrium temperatures below 400~K, as case studies and assess observations using different instrument configurations with one transit per instrument. We find that their JWST transmission spectra with modest observing time and adequate wavelength coverage can provide precise abundance constraints of key molecules H$_2$O, CH$_4$, and NH$_3$ even in the presence of clouds at significantly high altitudes. The best constraints are obtained by combining all three high-resolution NIRSpec gratings (G140H+G235H+G395H) that together span the $\sim$1-5~$\mu$m range. Single-transit observations with this three-instrument configuration allow precise abundance constraints for cloud-top pressures as low as 3~mbar and 0.1~mbar for K2-18~b and TOI-732~c, respectively, assuming a nominal 10$\times$~solar metallicity. The constraints vary with instrument combinations. We find that NIRSpec G235H+G395H is the optimal two-instrument configuration, while NIRISS or NIRSpec G235H is optimal for single-instrument observations. Absent high-altitude clouds, even single-instrument observations can provide good abundance constraints for these planets. Our findings underscore the promise of JWST transmission spectroscopy for characterising temperate mini-Neptunes orbiting nearby M dwarfs.
\end{abstract}




\section{Introduction} 
\label{sec:intro}
Our understanding of exoplanet atmospheres has improved dramatically over the last two decades. A wide range of spectroscopic observations have been used extensively to constrain atmospheric chemical compositions, temperature structures, clouds, hazes and other atmospheric properties \citep{Crossfield2015, Kreidberg2018, Madhu2019}. The method of transmission spectroscopy \citep{Charbonneau2002} has proven to be particularly effective in characterising atmospheric chemical compositions as well as clouds and hazes. The Hubble Space Telescope (HST) has been a key driver in this direction, starting with the detection of atomic species in the optical range using the Space Telescope Imaging Spectrograph (STIS) \citep{Charbonneau2002, VidalMadjar2003} and later the detections of H$_2$O with infrared observations from the Wide Field Camera 3 (WFC3) \citep{Deming2013, Huitson2013, Mccullough2014,Sing2016}. Additionally, advances in atmospheric modelling and retrievals led to the ability to place important constraints on atmospheric abundances and cloud/haze parameters for a number of giant exoplanets using such HST spectra \citep[e.g.][]{Madhu2014, Kreidberg2014,Barstow2017,Pinhas2019,Welbanks2019b}.

With the advent of the James Webb Space Telescope (JWST), the field will experience a generational shift in observational capabilities, starting from its very first observations \citep{Beichman2014, Stevenson2016, Batalha2017b, Bean2018, Kalirai2018, Sarkar2020}. Promising a virtually complete coverage of the near-infrared with better sensitivity than HST, JWST has much to offer. Notably, however, JWST's coverage does not include the full optical part of the spectrum. This makes it difficult to rely on optical observations to constrain the possible impact of clouds/hazes on the transmission spectrum, as has been pursued for hot Jupiters with HST \citep{ Sing2016, Barstow2017, Pinhas2019, Welbanks2019a}. 

One of the most exciting applications for JWST is in the characterisation of temperate mini-Neptunes, planets which have no solar system equivalent but are the most common type of exoplanet, along with super-Earths \citep{Fulton2017, Hardegree-Ullman2020}. Here we refer to mini-Neptunes as planets smaller than Neptune with volatile-rich interiors and H$_2$-rich atmospheres. Nominally, such planets could have radii in the range $\sim$1.6 - 4 R$_\oplus$, considering that most planets with radii larger than 1.6 R$_\oplus$ are unlikely to be rocky \citep{Rogers2015,Lozovsky2018}. One such example is K2-18~b \citep{Montet2015, Cloutier2017}, which has been shown with HST observations to contain H$_2$O in its hydrogen-dominated atmosphere \citep{Tsiaras2019, Benneke2019} and among the possible surface conditions is the existence of a habitable liquid ocean layer \citep{Madhusudhan2020}. For K2-18~b and many other such planets, more precise abundance estimates of key molecules, obtained from JWST observations, will lead to better constraints on surface conditions \citep{Hu2021a, Madhusudhan2021, Yu2021}. Additionally, the greater spectral coverage will help determine if CH$_4$ and NH$_3$, which are expected to be present from equilibrium calculations and so far have been undetected \citep{Benneke2019, Madhusudhan2020}, are indeed present and at what abundances. K2-18~b in particular is set to be observed as part of JWST Cycle 1 GO programs 2372 (PI: Renyu Hu) and 2722 (PI: Nikku Madhusudhan) for a total of 9 transits, the most of any mini-Neptune.

Given the multitude of instruments appropriate for transit spectroscopy aboard JWST, much effort has already been made to understand its observational capabilities. Several general studies \citep{Greene2016, Howe2017, Batalha2017a} considering a range of planets and employing various approaches to noise modelling and parameter estimation found that the $\sim$1-3$\mu$m range stands to be the most informative in constraining the atmosphere's metallicity. However, \citet{Guzman-Mesa2020}, using a random forest-based approach, find that for a relatively hot Neptune-like planet, the $\sim$3-5$\mu$m wavelength range is optimal instead, in the context of establishing atmospheric abundances and the C/O ratio. 

In addition to the above, there have been numerous studies specifically examining how JWST observations can be used to characterise super-Earths \citep[e.g.][]{Deming2009, Benneke2012, Molliere2017, Morley2017, Tremblay2020, Gialluca2021}. While JWST's generational improvements in wavelength coverage and resolution will be boons to super-Earth observation campaigns \citep{Tremblay2020, Gialluca2021}, they remain a sizeable undertaking, requiring  many transits to be observed for  their atmosphere to become detectable \citep{Barstow2016}. On the other hand, Mini-Neptunes, even temperate ones lying in or close to their host star's habitable zone, are significantly more amenable to spectroscopic characterisation than super-Earths, aided in no small part by their more extended, hydrogen-dominated atmospheres. 

Observing mini-Neptune atmospheres, however, is not without risks. Several previous attempts to constrain a mini-Neptune's atmospheric composition have been foiled by the possible presence of high-altitude clouds, most notably GJ~1214~b \citep{Bean2011,Desert2011,Kreidberg2014}. Depending on their altitude clouds can mask spectral signatures from molecules in the atmosphere, preventing their detection. Even if the molecular absorption features are not completely obscured by clouds, a lack of optical measurements could lead to potential degeneracies between any nominal cloud effects and the chemical abundances. 

Given the limited observing time available with JWST, the threat of clouds may scuttle attempts to fully explore the properties of temperate mini-Neptunes with JWST. Such a problem also exists for larger and hotter planets, and recent studies have explored the possibility of constraining atmospheric properties of  warm sub-Neptunes/Neptunes and hot Jupiters in the presence of clouds/hazes \citep[][]{Schlawin2018, Kawashima2019, Mai2019}. However, the problem is arguably more acute for temperate ($\lesssim$500~K) mini-Neptunes on which we focus here, whose lower temperatures decrease the extent of their atmospheres, resulting in weaker spectral signatures.  

In this work we present an optimal retrieval-driven approach to take advantage of JWST's observing capabilities to constrain atmospheric abundances of temperate mini-Neptunes in the presence of high-altitude clouds. We focus on temperate mini-Neptunes, as they are prime targets for JWST characterisation, with our findings also being relevant to larger and hotter planets offering an even better Signal-to-Noise Ratio (SNR). Using K2-18~b and TOI-732~c/LTT~3780~c as nominal examples, we generate cloudy and cloud-free synthetic transmission spectra and carry out atmospheric retrievals to determine how well atmospheric parameters can be constrained for different instrument configurations. We show that, thanks to the precision, spectral resolution and large wavelength coverage available with JWST, atmospheric retrievals can overcome high-altitude clouds and obtain precise abundance constraints for the dominant oxygen-, carbon- and nitrogen-carrying molecules in temperate mini-Neptunes. 

In Section \ref{sec:methods} we present our retrieval methodology, describing how we generate synthetic data as well as the retrieval approach used to analyse the data. Our results are presented in section \ref{sec:results}. We begin by considering a reference cloud-free atmosphere for the case of K2-18~b in section \ref{sec:cloud_free_reference}. We then investigate the more difficult cloudy case, again for K2-18~b in section \ref{sec:wavelength_coverage}, examining the performance of retrievals on observations combining one, two and three instruments with a single transit per instrument. Having established what abundance constraints are possible for K2-18~b, we then consider a more favourable target, TOI-732~c, as a case study in section \ref{sec:case_study}. We first establish how single-transit single-instrument observations of a cloudy atmosphere perform for this more favourable case, comparing our findings to those obtained for K2-18~b. We then benchmark the observing capabilities of JWST with a three-instrument configuration spanning the complete 1-5~$\mu$m range, seeking the highest altitude cloud deck with which molecular abundances can still be constrained for two atmospheric composition scenarios. Finally in section \ref{sec:conclusion}, we summarise and discuss our results about what can be expected from JWST  observations of temperate, cloudy mini-Neptunes.

\section{Methods}
\label{sec:methods}
Our goal is to determine how JWST observations can be used to successfully characterise mini-Neptune atmospheres despite the presence of high-altitude clouds. We explore a range of instrument configurations and atmospheric properties, seeking to establish what atmospheric parameter constraints can be obtained in each case. In this section, we present our approach.

\subsection{Case Studies and Canonical Atmospheric Model}
\label{sec:canonical_model}

\begin{table}
    \caption{ Planetary and stellar properties for the two case studies.}
    \centering
    \begin{tabular}{c|c|c}
       Properties  & K2-18~b & TOI-732~c \\
         \hline
        $R_{\mathrm{P}}$ / R$_{\earth}$                     &   2.61    &  2.42     \\
        $M_{\mathrm{P}}$ / M$_{\earth}$                     &   8.63    &  6.29     \\
        $g / \mathrm{m}\mathrm{s}^{-2}$      &   12.4    &  10.5     \\
        $T_{\mathrm{eq}}$ / K           &   282     &  363      \\
        $(\frac{R_\mathrm{P}}{R_\mathrm{\**}})^2$ (\%)     &  0.289     &   0.337    \\
        
        \multicolumn{3}{c}{\emph{Host Stars}}\\
       
        $T_{\mathrm{eff}}$ / K                              &   3503    &  3360     \\
        $R_{\mathrm{\**}}$ / R$_{\mathrm{\sun}}$            &   0.445   &  0.382    \\
        $M_{\mathrm{\**}}$ / M$_{\mathrm{\sun}}$            &   0.495   &  0.379    \\
        $J$ mag                                             &   9.8     &  9.0      \\
        \hline
               
    \end{tabular}
    \newline
    
    \footnotesize{Values for K2-18~b are from \citet{Cloutier2019} and \citet{Benneke2019}, while values for TOI-732~c are from \citet{Nowak2020}. $T_{\mathrm{eq}}$ is calculated assuming zero Bond albedo and full day-night redistribution.}
    \label{tab:planet_properties}
\end{table}

\begin{figure*}
    \includegraphics[width=\textwidth]{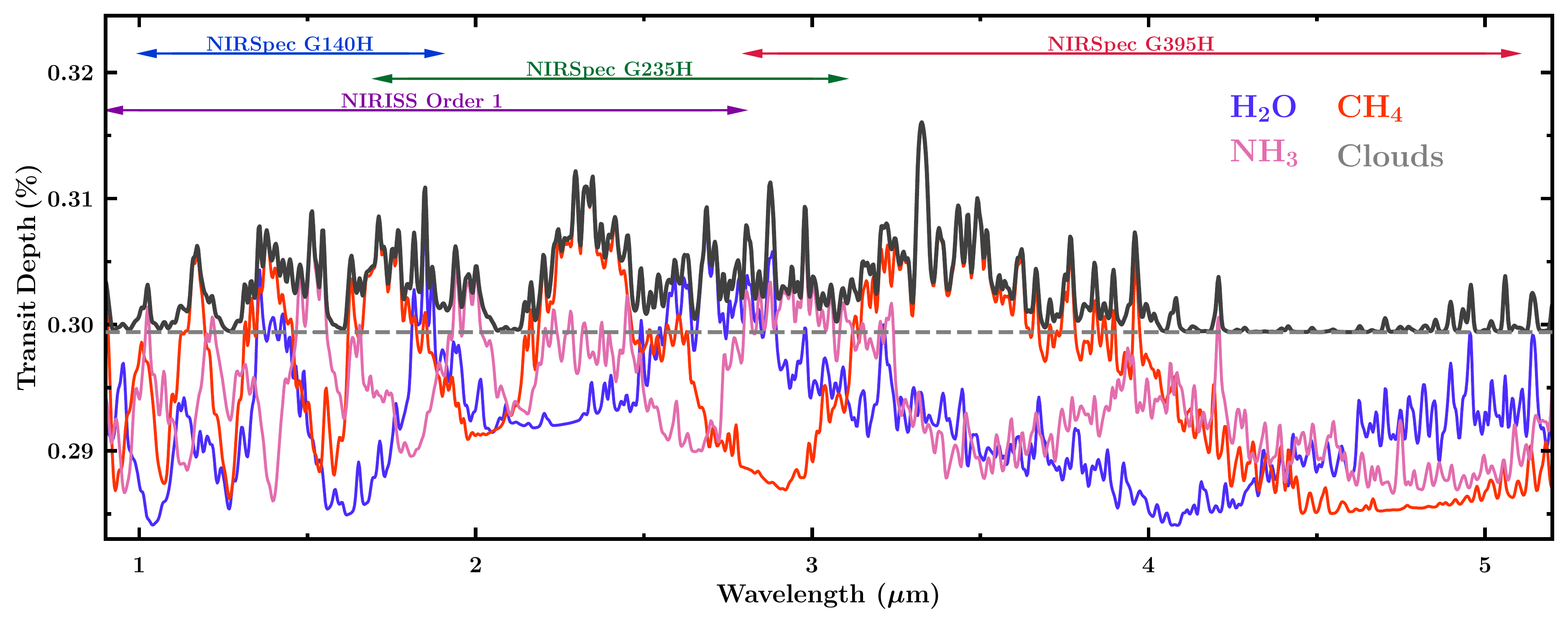}
    \caption{Contributions to a model transmission spectrum of K2-18~b corresponding to the canonical model described in section \ref{sec:canonical_model}. Each coloured line is the transmission spectrum produced exclusively by absorption from H$_2$O (blue), CH$_4$ (red) or NH$_3$ (pink). The grey dashed line shows the featureless spectrum with only the cloud deck present at 3~mbar, which sets the spectral baseline in the cloudy canonical model. The resulting transmission spectrum, with contributions from all chemical species and the cloud deck is shown in black. In all cases, we also include the effects of H$_2$-H$_2$ and H$_2$-He collision-induced absorption.}
    \label{fig:contribution_plot}
\end{figure*}

For our investigation, we consider two mini-Neptune planets as nominal examples, whose properties are summarised in table \ref{tab:planet_properties}. The first is K2-18~b \citep{Montet2015,Cloutier2017}, which we use for the majority of this work. Orbiting a J~=~9.8~mag M dwarf, K2-18~b has a mass of 8.63~$\pm$~1.35~M$_{\earth}$ \citep{Cloutier2019} and multiple radii reported in literature, with a value 2.51$^{+0.13}_{-0.18}$~R$_{\earth}$ obtained from K2 observations \citep{Hardegree-Ullman2020} and 2.610$\pm 0.087$~R$_{\earth}$ obtained using K2 and Spitzer observations \citep{Benneke2019}. Both are consistent to within 1-$\sigma$. For this work, we use a radius of 2.610$\pm 0.087$~R$_{\earth}$ \citep{Benneke2019}, for the sake of consistency with previous studies \citep[e.g.,][]{Madhusudhan2020}. We choose K2-18~b as a conservative case, with a low zero-albedo equilibrium temperature of 282~K and orbiting a star of intermediate brightness compared to more recently-discovered TOI targets such as TOI-175~d \citep{Kostov2019}, TOI-270~c and d \citep{Gunther2019, VanEylen2021}, TOI-732~c/LTT~3780~c \citep{Nowak2020, Cloutier2020} and TOI-776~b and c \citep{Luque2021}.

Our second case is the more recently-discovered TOI-732~c/LTT~3780~c \citep{Nowak2020, Cloutier2020}, with a mass of 6.29${^{+0.63}_{-0.61}}$~M$_{\earth}$ and radius of 2.42${^{+0.10}_{-0.10}}$~R$_{\earth}$, orbiting a relatively brighter J~=~9.0~mag host star \citep{Nowak2020}. It is also slightly hotter than K2-18~b, with an equilibrium temperature of 363~K. We choose TOI-732~c for our second case study as a more optimistic scenario, representative of the above mentioned population of mini-Neptunes that are more amenable to transmission spectroscopy, which JWST will likely be observing extensively over its lifetime.

For both planets, we consider a canonical model of a 1D plane-parallel atmosphere in hydrostatic equilibrium, dominated by H$_2$ and He in solar ratio. We additionally include H$_2$O, CH$_4$ and NH$_3$, the dominant O- C- and N-carrying molecules expected in temperate H$_2$-rich atmospheres in thermochemical equilibrium \citep{Burrows1999, Lodders2002, Madhu2011, Moses2013}. We use a nominal atmospheric composition arising from 10$\times$ solar elemental abundances \citep{Asplund2009} at chemical equilibrium. This corresponds to volume mixing ratios of $10^{-2}$, $5 \times 10^{-3}$ and $10^{-3}$ for H$_2$O, CH$_4$ and NH$_3$, respectively. We opt for 10$\times$~solar elemental abundances so that the H$_2$O mixing ratio matches the median H$_2$O estimate of $\sim10^{-2}$ obtained from HST observations of K2-18~b \citep{Benneke2019, Madhusudhan2020}. While CH$_4$ and NH$_3$ were not detected in K2-18~b, we use chemical equilibrium abundances rather than assuming any kind of depletion due to disequilibrium for this case study \citep[e.g.][]{Yu2021, Hu2021a, Hu2021b, Madhu2011}.

We assume an isothermal terminator atmospheric profile, in keeping with self-consistent P-T profile calculations in literature \citep{Benneke2019, Piette2020} finding that in the pressure range probed by transmission spectroscopy in K2-18~b the temperature structure is effectively isothermal. We set the terminator temperature profile to an isotherm at 300~K for K2-18~b, at the lower end of the temperate mini-Neptune range. Similarly to our choice of atmospheric composition, this choice is motivated by both the actual properties of K2-18~b (T$_{\mathrm{eq}}$~=~282~K), as well as seeking the most difficult case for retrievals, as a low temperature suppresses the atmospheric scale height. For TOI-732~c (T$_{\mathrm{eq}}$~=~363~K), we opt for an isotherm temperature of 350~K. While higher than what we use for K2-18~b, it is still relatively low, especially in comparison to many other similar mini-Neptunes, e.g. K2-3~b, TOI-270~c, etc. We nominally set the reference pressure ($P_{\rm ref}$) to be 0.1 bar for all the models in this work. 

For this canonical model, the H$_2$O spectral amplitude of a K2-18~b transmission spectrum in the 1.1-1.7 $\mu$m range is comparable to that observed in the HST WFC3 band \citep{Benneke2019, Madhusudhan2020}, modulo a constant offset in transit depth due to our choice of $P_{\rm ref}$. We also consider additional contributions from CH$_4$ and NH$_3$ in the models which are below the detection threshold with current HST data of K2-18~b. 

For TOI-732~c, we additionally consider a second atmospheric model, where H$_2$O is present at a volume mixing ratio of $10^{-2}$, the same as the canonical model, but CH$_4$ and NH$_3$ are depleted by 1 dex to volume mixing ratios of $5 \times 10^{-4}$ and $10^{-4}$, respectively. Such a composition therefore corresponds to either a solar metallicity atmosphere where H$_2$O has been enhanced, or a 10$\times$ solar metallicity atmosphere where CH$_4$ and NH$_3$ have been depleted. We do not include CO and CO$_2$ in our model atmospheres as their abundances in temperate H$_2$-rich atmospheres are expected to be very low under chemical equilibrium, compared to the prominent molecules we consider here. In principle, non-equilibrium mechanisms can enhance the amount of CO and CO$_2$ in the atmosphere to limits that may be detectable \citep{Hu2021a,Hu2021b,Yu2021}. We do not consider this aspect in the present study.

Some temperate mini-Neptunes could be effectively cloud-free in their observable atmospheres, while others may have high-altitude cloud decks. For the specific case of K2-18~b, \citet{Benneke2019} infer clouds in the lower atmosphere of K2-18~b using HST observations, with a model preference of 2.6~$\sigma$, and infer a cloud deck pressure range of 7.74–139~mbar with their retrievals and 10-1000~mbar using self-consistent atmospheric temperature models. \citet{Madhusudhan2020}, however, do not find strong evidence for clouds with the same data, reporting only a 1.1~$\sigma$ model preference.

We begin by considering the cloud-free case of K2-18~b as a reference. We then consider cases with clouds present, which we model as a grey opacity present at all pressures below the cloud-top pressure. We adopt a conservative cloud deck pressure of 3~mbar for our canonical model, motivated by both the findings of \citet{Benneke2019} mentioned above, as well as \citet{Blain2021} who found that optically-thick clouds in K2-18~b may form at pressures of 10~mbar or higher (i.e. deeper in the atmosphere), depending on metallicity. Our 3~mbar cloud deck therefore lies at a pressure $\sim$0.5~dex lower (i.e. at a higher altitude) than the 10~mbar theoretical estimates of \citet{Benneke2019} and \citet{Blain2021}. The spectral contributions of all three molecules as well as a 3~mbar cloud deck for K2-18~b are shown in figure \ref{fig:contribution_plot}. For TOI-732~c, we additionally consider even lower cloud deck pressures, seeking to benchmark the capabilities of JWST for the two atmospheric composition cases described above. We consider cloud deck pressures ranging from 100~bar, i.e., effectively cloud-free, to 0.03~mbar (3$\times 10^{-5}$~bar), i.e. high up in the atmosphere. For both planets, we use the worst-case scenario of 100$\%$ cloud coverage.

\subsection{Simulating JWST observations}
\label{sec:simulating_data}

\begin{figure*}
    \centering
    \includegraphics[width=\textwidth]{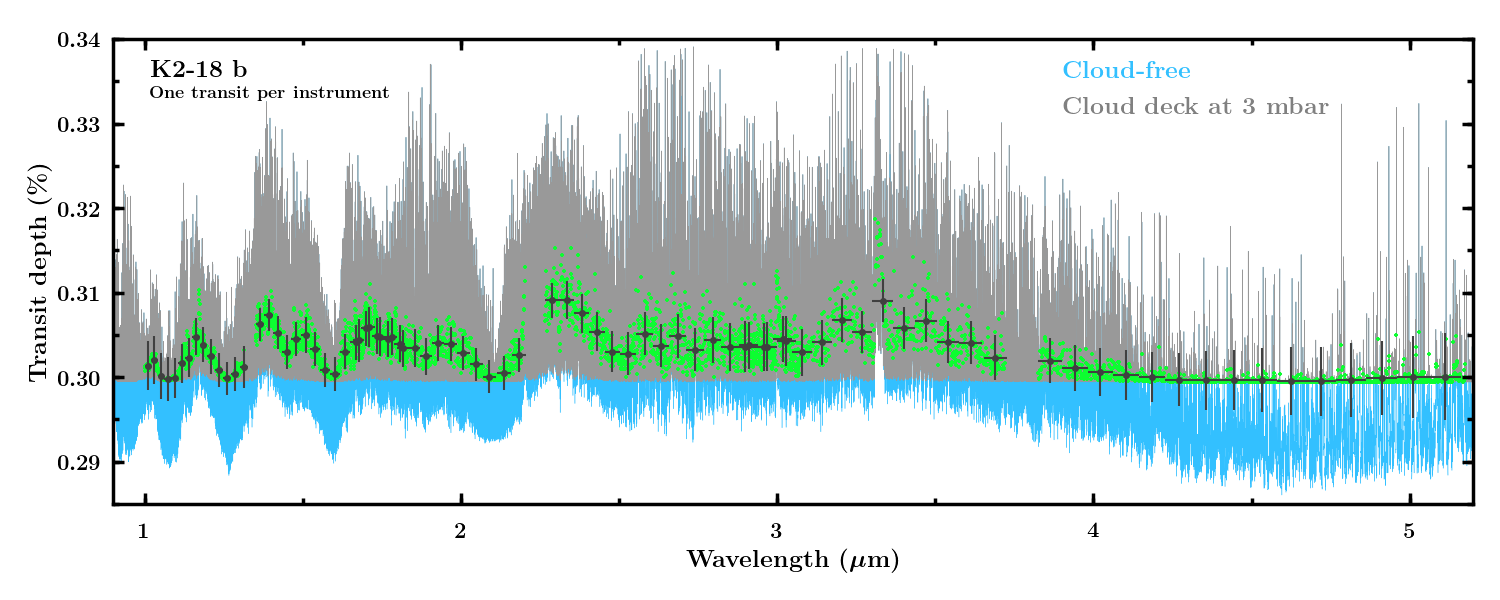}
    \caption{High resolution model transmission spectra of K2-18~b  corresponding to the canonical atmospheric model described in section \ref{sec:canonical_model} with a cloud deck at 3~mbar (grey) and no clouds (blue). Also shown are simulated observations generated from the cloudy model using NIRSpec gratings G140H, G235H and G395H at native resolution (green points) and binned to R=50 (black errorbars). For visual clarity the simulated data are centered on the binned model points.}
    \label{fig:K2-18b_data_generation}
\end{figure*}

We use the forward model-generating component of the AURA retrieval code \citep{Pinhas2018}, described in section \ref{sec:retrieval_methods} to generate a high-resolution model spectrum. This spectrum is input to the JWST simulator Pandexo \citep{Batalha2017b}, which we use to generate the wavelength bins and corresponding uncertainties for our simulated JWST observations. We do not explicitly set a noise floor, as the uncertainty per spectral element is never low enough in our case studies for it to be of concern. Throughout this work, we simulate the observation of one transit per instrument.

We consider simulated observations with the Near Infrared Spectrograph (NIRSpec) \citep{Ferruit2012, Birkmann2014} as well as the Near Infrared Imager and Slitless Spectrograph (NIRISS) \citep{Doyon2012}. For NIRSpec, we consider all three high-resolution gratings: G140H, G235H and G395H, paired with the F100LP, F170LP and F290LP filters, respectively. With this configuration, observations are obtained over a wavelength range of 1.0-1.8~$\mu$m for G140H, 1.7-3.1~$\mu$m for G235H and 2.9-5.1~$\mu$m for G395H. For all three gratings we use the SUB2048 subarray for maximal wavelength coverage and the NRSRAPID readout mode. The three gratings achieve a resolution of R~$\sim$2700. Additionally, we use NIRISS in Single Object Slitless Spectroscopy mode (SOSS), with the GR700XD grism, SUBSTRIP96 and NISRAPID readout mode , yielding observations with a wavelength coverage of 0.9-2.8~$\mu$m and a resolution of R~$\sim$700 with order 1.

We do not consider observations with the Near Infrared Camera as in its Wide Field Slitless Spectroscopy mode it offers lower wavelength coverage and resolution than those provided by the high-resolution NIRSpec gratings. We also do not consider NIRSpec PRISM in our investigations, despite its large wavelength coverage, due to its faint saturation threshold (J $\gtrsim$ 10.5). Lastly, we do not consider pairing the NIRSpec G140H grating with the F070LP filter, as this achieves a comparatively small wavelength coverage between 0.8-1.3~$\mu$m, which largely overlaps with that obtained with G140H/F100LP. 

Throughout this work, we conduct retrievals on all simulated observations at their native resolution, rather than binning down to a particular resolution. We find that retrievals on native resolution observations achieve more accurate and robust abundance estimates, especially in the case of very high altitude cloud decks as discussed in section~\ref{sec:benchmark_canonical}. Figure \ref{fig:contribution_plot} shows the wavelength coverage of NIRISS and the three NIRSpec gratings, as well as the spectral contributions from the three molecules considered in this work, H$_2$O, CH$_4$ and NH$_3$. The masking effect of clouds at 3~mbar is shown as a grey dashed line. 

To generate simulated JWST observations, we first convolve the model high-resolution spectrum with each instrument's spectral point-spread function \citep{Perrin2012, Sarkar2021}. We then bin the convolved spectrum down to the Pandexo-generated wavelength bins, taking into account each instrument's sensitivity function. We lastly add Gaussian noise to the spectrum, offsetting each datapoint according to its Pandexo-provided uncertainty. As this step may introduce artefacts in the data that are specific to a particular noise instance, we carry out several retrievals for each case presented, re-generating the data with a different noise instance each time. Figure \ref{fig:K2-18b_data_generation} shows such simulated observations for K2-18~b with our canonical cloudy model, simulated for the three high-resolution NIRSpec gratings at native resolution and at R~=~50. The data shown has no noise added for visual clarity. Also shown are the cloudy and cloud-free high-resolution transmission spectra we use to generate synthetic observations.

\subsection{Retrieval Methodology}
\label{sec:retrieval_methods}

For this work we use the latest version of the AURA retrieval code \citep{Pinhas2018}. AURA comprises of two parts. The first is a forward model generator, which was also used in creating the simulated JWST observations described in section \ref{sec:simulating_data}. The second is a robust Bayesian parameter estimator based on the PyMultiNest Nested Sampling package \citep{Buchner2014}.

Our forward model generator computes radiative transfer for a 1D plane-parallel atmosphere under hydrostatic equilibrium to create transmission spectra. The atmosphere's chemical composition is treated as uniform in altitude, with the mixing ratio of each constituent chemical species being a free parameter, avoiding any assumptions of equilibrium chemistry. As mentioned in section \ref{sec:canonical_model}, we model the temperature profile as an isotherm whose temperature is a single free parameter.

We include opacity contributions from H$_2$O \citep{Rothman2010}, CH$_4$ \citep{Yurchenko2014} and NH$_3$ \citep{Yurchenko2011}, as well those arising from H$_2$-H$_2$ and H$_2$-He collision-induced absorption (CIA) \citep{Richard2012}. These are included as they are the most dominant species expected in thermochemical equilibrium in temperate H$_2$-rich atmospheres \citep{Burrows1999, Madhu2011, Moses2013}. As discussed in section \ref{sec:canonical_model}, we do not include less abundant species such as CO or CO$_2$ in our retrievals. Clouds are treated as a grey opacity across the entire wavelength range, described by the pressure at the top of the cloud deck, as well as the fraction of the terminator atmosphere covered by clouds. In total, our atmospheric model has 7 free parameters: 3 of which are the log-mixing ratios of H$_2$O, CH$_4$ and NH$_3$, while the remaining four are the isotherm temperature $T_{\mathrm{iso}}$, the reference pressure, $P_{\mathrm{ref}}$, corresponding to the pressure at the given planet's radius, the cloud deck pressure, $P_{\mathrm{c}}$ and fractional cloud coverage ${\bar \phi}$.

 For our retrievals, we use log-uniform volume mixing ratio priors between 10$^{-12}$-10$^{-0.3}$ for all three chemical species. For the isotherm temperature, we use a relatively uninformative uniform prior between 50-600~K. We note that while retrievals on actual data will most likely use a smaller prior range whose upper limit will be informed by the planet's equilibrium temperature, we do not do so here so as to fully capture any potential degeneracies involving atmospheric temperature. We note that for each forward model generated by our retrieval, the atmospheric scale height is recomputed for its specific mean molecular weight and atmospheric temperature. For both the reference pressure and cloud deck pressure, we use priors that span the full extent of our model atmosphere. We additionally use a uniform prior for the cloud coverage fraction, ranging from 0 to 1.

\section{Results}
\label{sec:results}

In this section we present our findings on how JWST observations can provide abundance constraints for temperate mini-Neptune atmospheres in the presence of high-altitude clouds. As mentioned in section \ref{sec:methods}, we consider single, double and triple instrument configurations consisting of NIRISS, covering the 0.9-2.8~$\mu$m wavelength range, as well as NIRSpec G140H, G235H and G395H configurations spanning wavelength ranges of 1.0-1.8, 1.7-3.1 and 2.9-5.1~$\mu$m, respectively. We simulate observing only one transit with each instrument considered.

We first examine the cloud-free case for K2-18~b in section \ref{sec:cloud_free_reference}, establishing a reference against which results from retrievals on cloudy atmospheres will be compared. We then examine the effects of instrument choice and wavelength coverage for retrievals on a cloudy atmosphere in section \ref{sec:wavelength_coverage}, again for the case of K2-18~b. We subsequently present retrieval results for TOI-732~c in section \ref{sec:case_study}, a more spectroscopically amenable target, starting with the minimal single-transit single-instrument setup in section \ref{sec:TOI-732_single_instruments}. We then consider observations combining all three NIRSpec gratings for atmospheric conditions ranging from cloud free to the highest-altitude cloud deck with which molecular abundances can still be constrained, thereby benchmarking what is achievable with JWST observations. We carry this investigation out for both the canonical model composition used for K2-18~b in section \ref{sec:benchmark_canonical}, as well as a second case, where the abundances of CH$_4$ and NH$_3$ are reduced by 1~dex relative to the canonical model in section \ref{sec:benchmark_depleted}.

\begin{table}
    \centering
    \caption{ Retrieved log-mixing ratio constraints for K2-18~b for all atmospheric and instrumental configurations considered.}
    \begin{tabular}{l|c|c|c}
        & \multicolumn{3}{c}{log-Mixing Ratios} \\[0.5mm]
        Case &  H$_2$O & CH$_4$ & NH$_3$  \\[0.5mm]
        \hline
        \hline
        True Values& -2 & -2.3 & -3 \\[0.5mm]
        \hline
        \multicolumn{4}{c}{Cloud-Free} \\[0.5mm]
        \hline
        \multicolumn{4}{l}{\emph{Single-Instrument Configurations}} \\[0.5mm]
        NIRSpec G140H                   &   $(-1.80)$                   &   $-2.69^{+0.87}_{-0.65}$    &   $-4.37^{+0.93}_{-0.73}$  \\[0.5mm]
        NIRSpec G235H                   &   $-1.73^{+0.53}_{-0.82}$     &   $-2.34^{+0.37}_{-0.49}$    &   $-2.78^{+0.39}_{-0.49}$  \\[0.5mm]
        NIRSpec G395H                   &   $(-2.18)$                     &   $-3.06^{+1.00}_{-1.04}$    &   $-3.58^{+1.08}_{-1.04}$  \\[0.5mm]
        NIRISS                          &   $-2.06^{+0.54}_{-0.72}$     &   $-2.19^{+0.47}_{-0.46}$    &   $-3.30^{+0.48}_{-0.51}$  \\[0.5mm]
        \\[0.5mm]
        \multicolumn{4}{l}{\emph{Three-Instrument Configuration}} \\[0.5mm]
        G140H + G235H + G395H           &   $-1.93^{+0.27}_{-0.36}$     &   $-2.34^{+0.24}_{-0.26}$    &   $-3.01^{+0.24}_{-0.27}$  \\[0.5mm]
        \hline
        \multicolumn{4}{c}{3~mbar Cloud Deck} \\[0.5mm]
        \hline
        \multicolumn{4}{l}{\emph{Single-Instrument Configurations}} \\[0.5mm]
        NIRSpec G140H                   &   $(-1.93)$                     &   $-3.85^{+1.65}_{-0.87}$    &   $-4.70^{+1.71}_{-0.89}$  \\[0.5mm]
        NIRSpec G235H                   &   $(-1.53)$                     &   $-1.51^{+0.75}_{-1.04}$    &   $-2.63^{+0.97}_{-1.46}$  \\[0.5mm]
        NIRSpec G395H                   &   $(-2.47)$                     &   $-4.81^{+2.39}_{-0.96}$    &   $(-1.58)$  \\[0.5mm]
        NIRISS                          &   $(-2.95)$                     &   $-3.38^{+1.20}_{-0.84}$    &   $(-2.02)$     \\[0.5mm]
        \\[0.5mm]
        \multicolumn{4}{l}{\emph{Two-Instrument Configurations}} \\[0.5mm]
        NIRSpec G140H + G235H           &   $-2.72^{+0.79}_{-0.85}$     &   $-2.45^{+0.49}_{-0.48}$    &   $-3.30^{+0.55}_{-0.55}$  \\[0.5mm]
        NIRSpec G140H + G395H           &   $(-1.51)$                     &   $-2.51^{+0.98}_{-1.08}$    &   $-3.10^{+0.99}_{-1.16}$  \\[0.5mm]
        NIRSpec G235H + G395H           &   $-2.48^{+0.87}_{-0.89}$     &   $-2.67^{+0.59}_{-0.55}$    &   $-3.18^{+0.60}_{-0.57}$  \\[0.5mm]
        NIRISS + NIRSpec G395H          &   $-2.86^{+0.88}_{-1.60}$    &    $-2.37^{+0.56}_{-0.58}$    &   $-3.12^{-0.60}_{-0.62}$  \\[0.5mm]
        \\[0.5mm]
        \multicolumn{4}{l}{\emph{Three-Instrument Configuration}} \\[0.5mm]
        G140H + G235H + G395H           &   $-2.22^{+0.55}_{-0.77}$    &    $-2.51^{+0.45}_{-0.52}$    &   $-3.16^{+0.48}_{-0.58}$     \\[0.5mm]
        \hline

    \end{tabular}
    \newline
    \footnotesize{Note: In cases where the lower 1-$\sigma$ interval spans more than 2 dex, we instead list the 2-$\sigma$ (95\%) upper estimate in brackets.}
    \label{tab:K2-18b_results}
\end{table}

\subsection{A Reference Cloud-free Case of K2-18~b}
\label{sec:cloud_free_reference}

The best-case scenario for transmission spectroscopy is when the planet's cloud deck lies below the observable slant photosphere i.e. at altitudes that do not contribute to its transmission spectrum. Such atmospheres are therefore effectively "cloud-free" from a spectroscopic characterisation standpoint. As a result, molecular absorption features are not masked by a cloud deck's opacity contributions, illustrated in figures \ref{fig:contribution_plot} and \ref{fig:K2-18b_data_generation}, making such spectra most favourable for constraining atmospheric abundances and other properties.

We begin by considering such a cloud-free case for K2-18~b, establishing a reference for what is achievable without the truncating effect of clouds. In this section we first consider observations spanning the $\sim$1-5~$\mu$m range by combining NIRSpec G140H, G235H and G395H. We then consider single-instrument observations with the three NIRSpec gratings as well as NIRISS. As in the rest of this work, we simulate {the observation of a single transit} per instrument. The retrieved abundance constraints are summarised in table \ref{tab:K2-18b_results}.

\begin{figure*}
    \centering
    \includegraphics[width=0.995\textwidth]{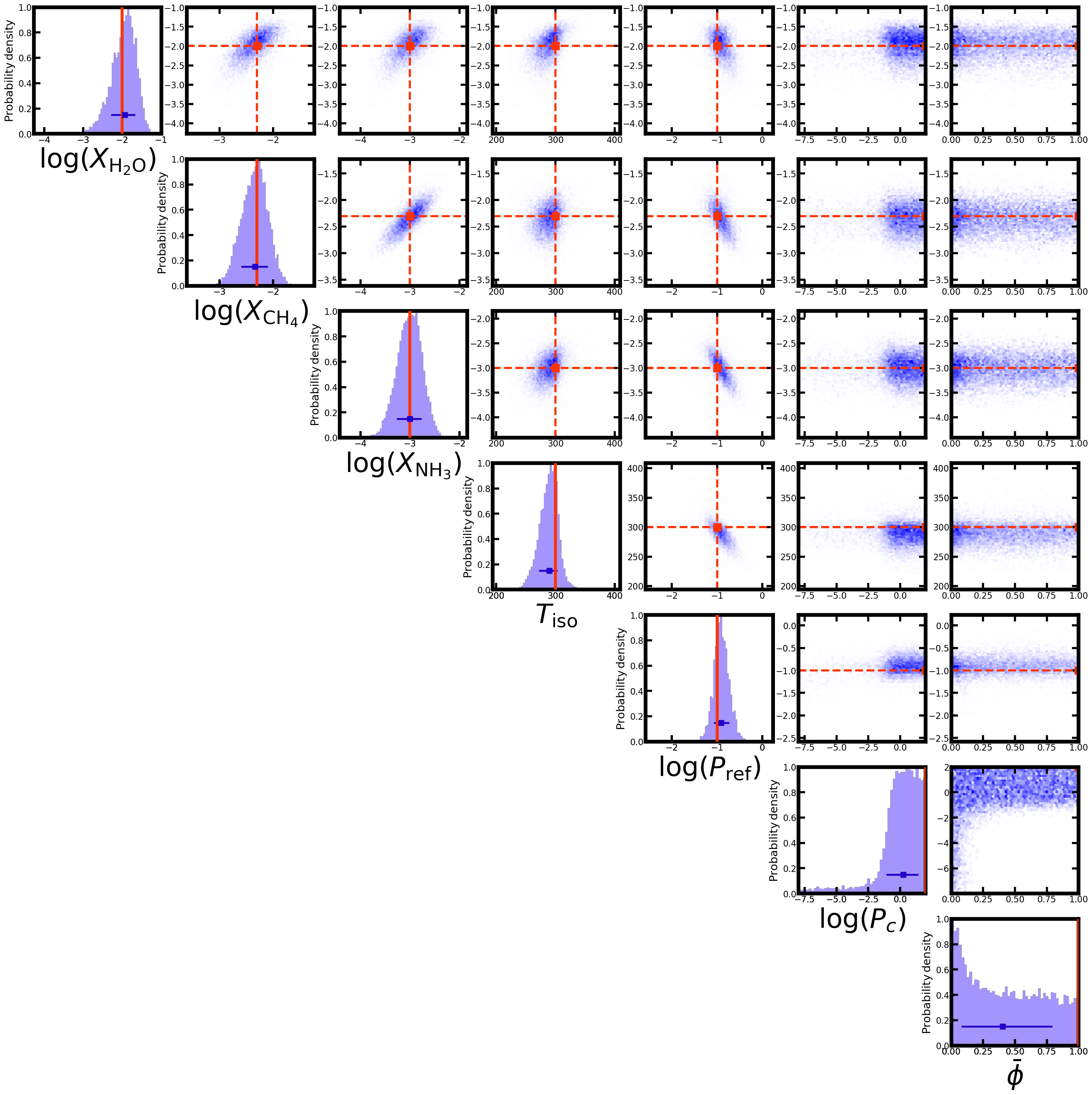}
    \caption{Posterior probability distribution for all 7 retrieved parameters obtained by a retrieval on combined NIRSpec G140H, G235H and G395H simulated observations of cloud-free K2-18~b. The data were generated for a cloud-free atmosphere with a composition corresponsing to 10$\times$ solar elemental abundances, as described in section \ref{sec:canonical_model}. The blue squares with error bars denote the median and 1-$\sigma$ intervals, while red solid and dashed lines denote the true values used to generate the data.}
    \label{fig:threeinstrument_clear_corner}
\end{figure*}

\subsubsection{Three-Instrument Observations}

 We first consider an instrument configuration combining observations from NIRSpec G140H, G235H and G395H. Figure \ref{fig:threeinstrument_clear_corner} shows the posterior probability distribution for all 7 retrieved parameters and the retrieved mixing ratio constraints are summarised in table \ref{tab:K2-18b_results}. The retrieval obtains log-mixing ratio estimates of $-1.93^{+0.27}_{-0.36}$ for H$_2$O, $-2.34^{+0.24}_{-0.26}$ for CH$_4$ and $-3.01^{+0.24}_{-0.27}$ for NH$_3$.  It can be seen that the retrieval constrains the cloud deck pressure, $P_{\mathrm{c}}$, to high values. While there is some spread towards lower pressures, the correlation plot of $P_{\mathrm{c}}$ with the cloud fraction, $\bar \phi$, shows that they are associated with cloud fractions near zero and thus have minimal impact on the spectrum.

 Our retrievals achieve abundance constraints mostly below $\sim$0.3 dex with only 3 transits observed in total. Moreover, such abundance constraints have been achieved without any supporting observations in the optical to constrain the spectrum's baseline. As can be seen in figure \ref{fig:contribution_plot}, the large wavelength coverage JWST offers encompasses numerous absorption peaks from all three species present in our models. This enables retrievals to use the relative heights of several peaks and the depths of troughs between them to implicitly ascertain the spectrum's baseline.
 
 We conduct repeat retrievals, each time generating the synthetic data anew with a different noise instance, to ensure our results are not caused by noise-specific features. We find that the majority of retrievals find abundance estimates that are within 1-$\sigma$ of the true values while a minority obtain abundance estimates that are between 1 and 2 $\sigma$ away from the true values.

\begin{figure*}
    \centering
    \includegraphics[width=\textwidth]{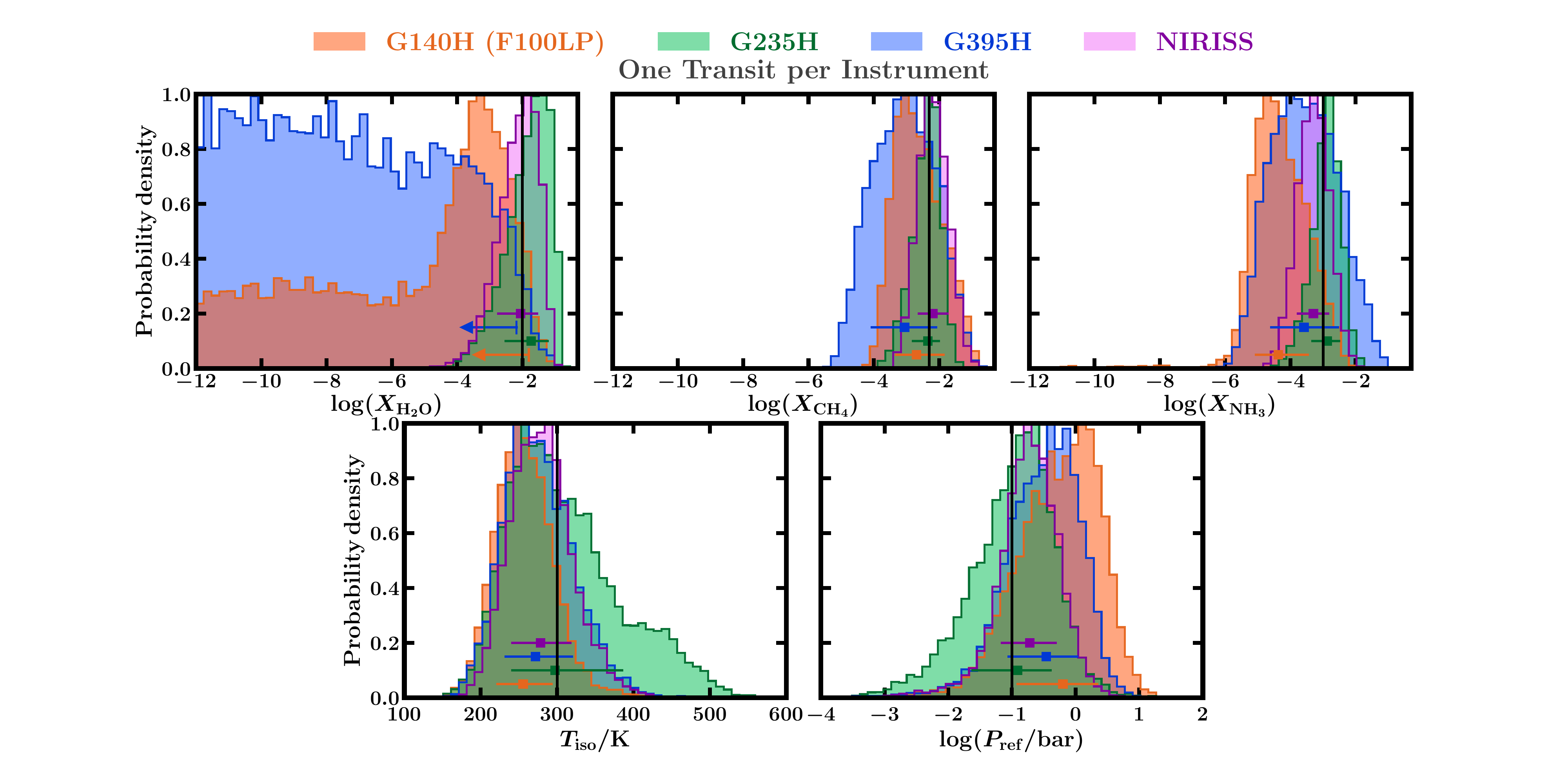}
    \caption{Posterior distributions of retrievals carried out on simulated single-instrument observations of K2-18~b without high-altitude clouds, using NIRSpec G140H (orange), G235H (green) and G395H (blue) as well as NIRISS order 1 (purple), retrieving on data from one instrument at a time. All data were generated for a nominal atmosphere at 10$\times$ solar elemental abundances, as described in section \ref{sec:canonical_model}. Black vertical lines denote the true values used to generate the data, while horizontal errorbars denote the median value and 1-$\sigma$ intervals. If a retrieval produces a lower 1-$\sigma$ interval spanning more than 2~dex, the errorbar is replaced with an arrow denoting the 2-$\sigma$ (95\%) confidence upper limit.}
    \label{fig:singleinstrument_clear_posteriors}
\end{figure*}

\subsubsection{Single-Instrument Observations}

Figure \ref{fig:singleinstrument_clear_posteriors} shows posterior distributions obtained from retrievals conducted on single-instrument observations, again without clouds present. The substantial decrease of wavelength coverage compared to the full $\sim$1-5~$\mu$m coverage used previously leads to less precise parameter constraints and even two cases where the mixing ratio of H$_2$O is largely unconstrained. Our findings indicate that observations solely from either NIRISS or NIRSpec G235H can precisely retrieve the correct abundances for H$_2$O, CH$_4$ and NH$_3$. In the case of NIRISS, we obtain an H$_2$O mixing ratio estimate with 0.5 and 0.7~dex upper and lower 1-$\sigma$ intervals, respectively. CH$_4$ and NH$_3$ are retrieved even more precisely, with a 1-$\sigma$ uncertainty of 0.5~dex for both molecules. Using NIRSpec G235H, we obtain constraints comparable to those obtained with NIRISS, finding upper and lower 1-$\sigma$ intervals of 0.5 and 0.8~dex, respectively, for H$_2$O and 0.4 and 0.5~dex for both CH$_4$ and NH$_3$. As can be seen in figure \ref{fig:singleinstrument_clear_posteriors}, the retrieval on NIRSpec G235H data achieved comparable abundance constraints, despite yielding a less precise estimate for the terminator temperature and reference pressure.

With single-transit NIRSpec G140H observations, we find that the abundances of CH$_4$ and NH$_3$ are still successfully constrained, but somewhat less precisely. Specifically, our retrieval produces upper and lower 1-$\sigma$ uncertainties of 0.9 and 0.7~dex for both CH$_4$ and NH$_3$. The NH$_3$ median estimate lies slightly more than 1-$\sigma$ away from the true value, but is still well within 2-$\sigma$. The H$_2$O posterior, however, shows a peak but it is not constrained, indicating that while H$_2$O is potentially present in the atmosphere, only an upper limit can be placed on its mixing ratio.

NIRSpec G395H observations also lead to constraints for the mixing ratios of CH$_4$ and NH$_3$ but not for H$_2$O. The retrieval produces mixing ratio estimates for CH$_4$ and NH$_3$ with uncertainties larger than those obtained with NIRSpec G140H, at $\sim$1.0 and $\sim$1.1~dex, respectively. As seen in figures \ref{fig:contribution_plot}, the NIRSpec G395H wavelength range contains significant spectral features from all three molecules in our model. However, the comparatively lower SNR that NIRSpec G395H achieves means that observing a single transit is insufficient to constrain all three molecules for our limiting case of K2-18~b.

As before, we conduct repeat retrievals on synthetic data with different noise instances to ensure the robustness of our findings. In the case of G140H, we find that subsequent retrievals sometimes under-estimate the abundances of CH$_4$ and NH$_3$, offering median values that are slightly more than 1-$\sigma$ below from the true value. Some G395H retrievals yield an H$_2$O posterior which is peaked at the true value, but nevertheless are poorly constrained and display a sizeable ``tail'' towards lower abundances. Retrievals on G235H and NIRISS show little variability, consistently yielding abundance estimates that are within 1-$\sigma$ of the true values most times and occasionally between 1- and 2-$\sigma$, as expected.

The generally less precise constraints of our single-transit NIRSpec G140H and G395H retrievals indicate that for observations of planets with comparable precision to K2-18~b and with atmospheric properties similar to our canonical model, multiple transits may have to be observed, or be combined with observations from NIRISS or NIRSpec G235H. On the other hand, both NIRSpec G235H and NIRISS consistently produce abundance constraints of comparable precision to those shown in table \ref{tab:K2-18b_results}, thanks to both covering a feature-rich part of the IR spectrum. This is in spite of NIRSpec G235H offering a smaller wavelength coverage than NIRISS, as it achieves a higher precision (accounting for resolution) as well as resolution, which is enough to compensate for its narrower bandpass.

From our investigation of cloud-free K2-18~b, we find that JWST observations, even those from a single instrument such as NIRISS or NIRSpec G235H observing a single transit, can lead to precise abundance constraints. Moreover, this is done without any supporting optical data. In the following section, we examine whether the same is true when high-altitude clouds obscure part of the planet's transmission spectrum.

\subsection{Case Study: K2-18~b with Clouds}
\label{sec:wavelength_coverage}

High-altitude clouds pose a significant challenge for retrievals. Opacity contributions from clouds mask spectral features, resulting in shallower, truncated troughs between absorption peaks. Consequently, retrievals may be unable to distinguish between features that are masked by clouds and those that are merely the result of a smaller scale height or low abundances. This can manifest as poorly constrained posterior distributions, or even as doubly-peaked posteriors, indicating two competing explanations for the data between which the retrieval is unable to distinguish.

We now investigate abundance constraints for K2-18~b in the presence of a high-altitude grey opacity cloud deck at 3~mbar, motivated by theoretical and observational grounds, as discussed in section \ref{sec:canonical_model}. We systematically investigate whether instrument combinations up to the three-instrument NIRSpec G140H+G235H+G395H combination, used in Section \ref{sec:cloud_free_reference}, can successfully constrain atmospheric properties using different combinations of wavelength coverage. Using the same nominal K2-18~b model as above, we examine all possible non-overlapping combinations of observations with NIRSpec G140H (F100LP filter), G235H and G395H as well as NIRISS. The precise configurations used to simulate each instrument are detailed in Section \ref{sec:simulating_data}. As before, we simulate observing a single transit with each instrument.

We first consider observations with individual instruments in section \ref{sec:single_transit_obs}, as was done for the cloud-free case in section \ref{sec:cloud_free_reference}, before moving on to two-instrument configurations in section \ref{sec:two_transit_obs}. Since we seek to ascertain the effect of wavelength coverage on retrieved parameters, we do not consider two-instrument cases where both largely cover the same wavelength range, i.e. NIRISS + NIRspec G140H or NIRISS + NIRspec G235H. For the same reason, we do not consider three-instrument configurations other than NIRSpec G140H + G235H + G395H presented in section \ref{sec:three_instrument_obs} All retrieved abundance constraints are summarised in table \ref{tab:K2-18b_results}, in addition to those obtained above for the cloud-free case.

\subsubsection{Single-Instrument Observations}
\label{sec:single_transit_obs}

\begin{figure*}
    \centering
    \includegraphics[width=\textwidth]{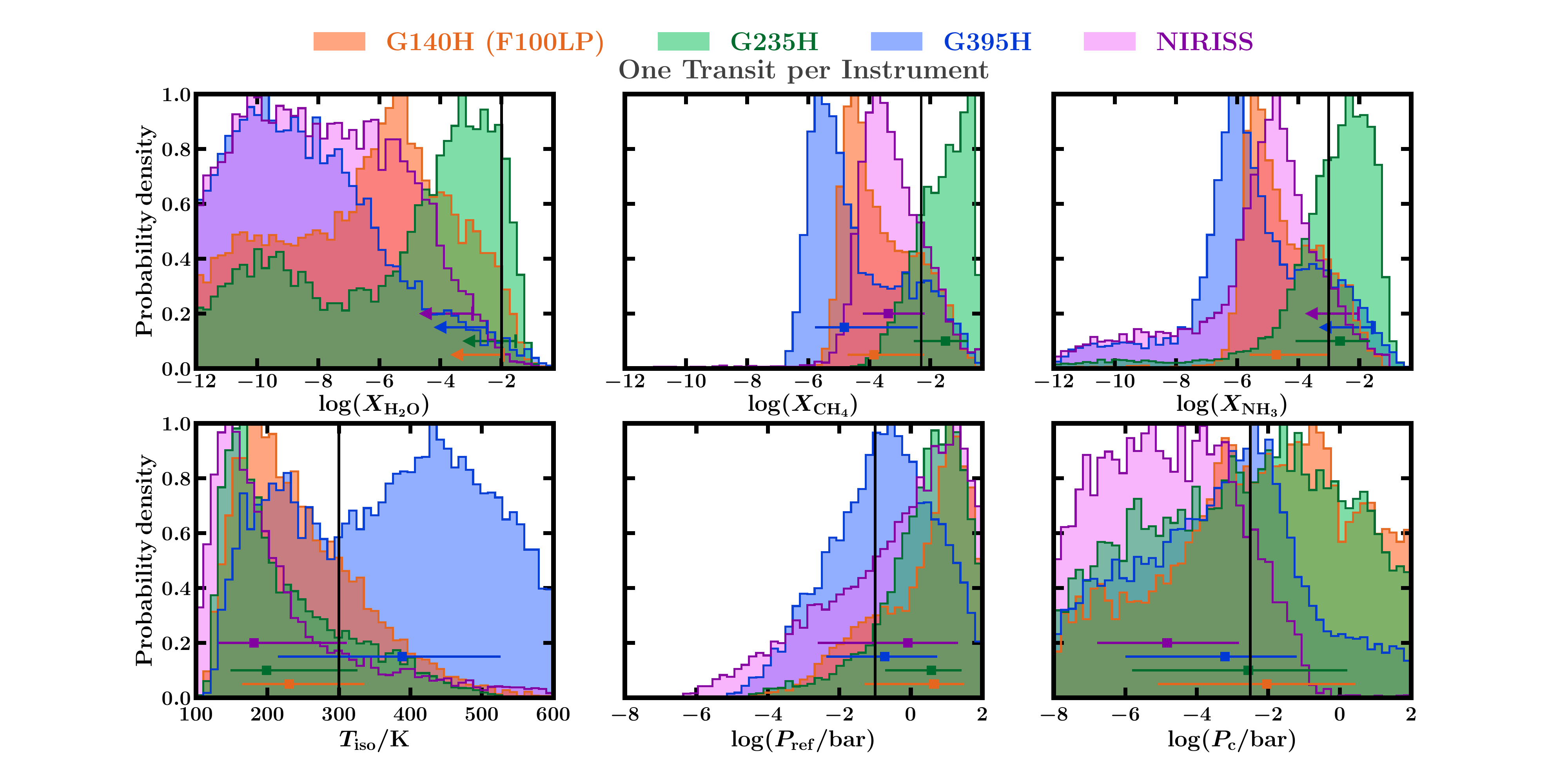}
    \caption{Posterior distributions for retrievals carried out on simulated single-instrument observations for the conservative case of K2-18~b with clouds, using NIRSpec G140H (orange), G235H (green) and G395H (blue) as well as NIRISS (purple), retrieving on data from one instrument at a time. All data were generated for a nominal atmosphere at 10$\times$ solar elemental abundance with a cloud deck at 3~mbar. Black vertical lines denote the true values used to generate the data, while horizontal errorbars denote the median value and 1-$\sigma$ intervals. If a  retrieval results in a lower 1-$\sigma$ uncertainty larger than 2~dex, the errorbar is replaced with an arrow denoting the 2-$\sigma$ (95\%) confidence upper limit.}
    \label{fig:singleinstrument_clouds_posteriors}
\end{figure*}

We first consider single-instrument observations with a cloud deck at a pressure of 3~mbar. As seen in the posterior distributions shown in Figure \ref{fig:singleinstrument_clouds_posteriors}, clouds cause a significant deterioration to the parameter constraints obtained. Three of the four retrievals are unable to meaningfully constrain the cloud deck pressure, while the retrieval on NIRISS data somewhat constrains the cloud deck pressure to lower values.  As a result, poor constraints are also obtained for the other parameters, as the retrievals are unable to ascertain if the spectral features are truncated by clouds or were instead produced by a small scale height or low abundances.

The retrieval on NIRSpec G140H observations produces estimates for the mixing ratios of NH$_3$ and CH$_4$ that are within 1-$\sigma$ of the true values, with both having upper and lower 1-$\sigma$ intervals of +1.7 and -0.9 dex. The retrieval only yields an upper bound for the mixing ratio of H$_2$O. The findings of the retrieval on NIRSpec G235H are similar to what was obtained by the retrieval on NIRSpec G140H data: H$_2$O is again largely unconstrained but its posterior distribution is more significantly peaked near the true value, while CH$_4$ and NH$_3$ mixing ratios are estimated accurately. CH$_4$ was retrieved with upper and lower 1-$\sigma$ intervals of +0.8 and -1.0~dex, respectively, while NH$_3$ was constrained with upper and lower 1-$\sigma$ intervals of +1.0 and -1.5~dex, respectively.

Retrieving on NIRISS data resulting in the log-mixing ratio of CH$_4$ being constrained with upper and lower 1-$\sigma$ intervals of 1.2 and 0.8~dex, respectively. The retrieval was only able to produce upper bounds for the mixing ratios of H$_2$O and NH$_3$. This is due to NIRISS achieving a lower spectroscopic precision (accounting for resolution) as well as a lower resolution than the high-resolution NIRSpec gratings, despite offering the largest wavelength coverage of the four instruments considered. Its larger wavelength coverage does not compensate for its lower precision, unlike in the cloud-free case. Due to its larger wavelength coverage, however, it is the only retrieval that led to a tentative indication that clouds may be present in the planet's atmosphere, somewhat constraining $P_\mathrm{c}$ to lower values that have an impact on the transmission spectrum. It is therefore possible to conclude that the the poorer constraints relative to the cloud-free case are most likely caused by high-altitude clouds.

The retrieval on NIRSpec G395H also successfully detects CH$_4$, producing a mixing ratio estimate with upper and lower 1-$\sigma$ intervals of 2.4 and 1.0~dex, respectively. It also indicates that NH$_3$ may potentially be also present, but the notable spread shown by its posterior distribution towards low abundances implies that models where NH$_3$ does not significantly contribute to the spectrum are also plausible. Additionally, H$_2$O is unconstrained. Notably, the retrieval produced doubly-peaked posterior distributions for the mixing ratios of CH$_4$ and NH$_3$ as well as $T_\mathrm{iso}$, indicating that there are two competing explanations for the data.

\begin{figure*}
    \centering
    \includegraphics[width=\textwidth]{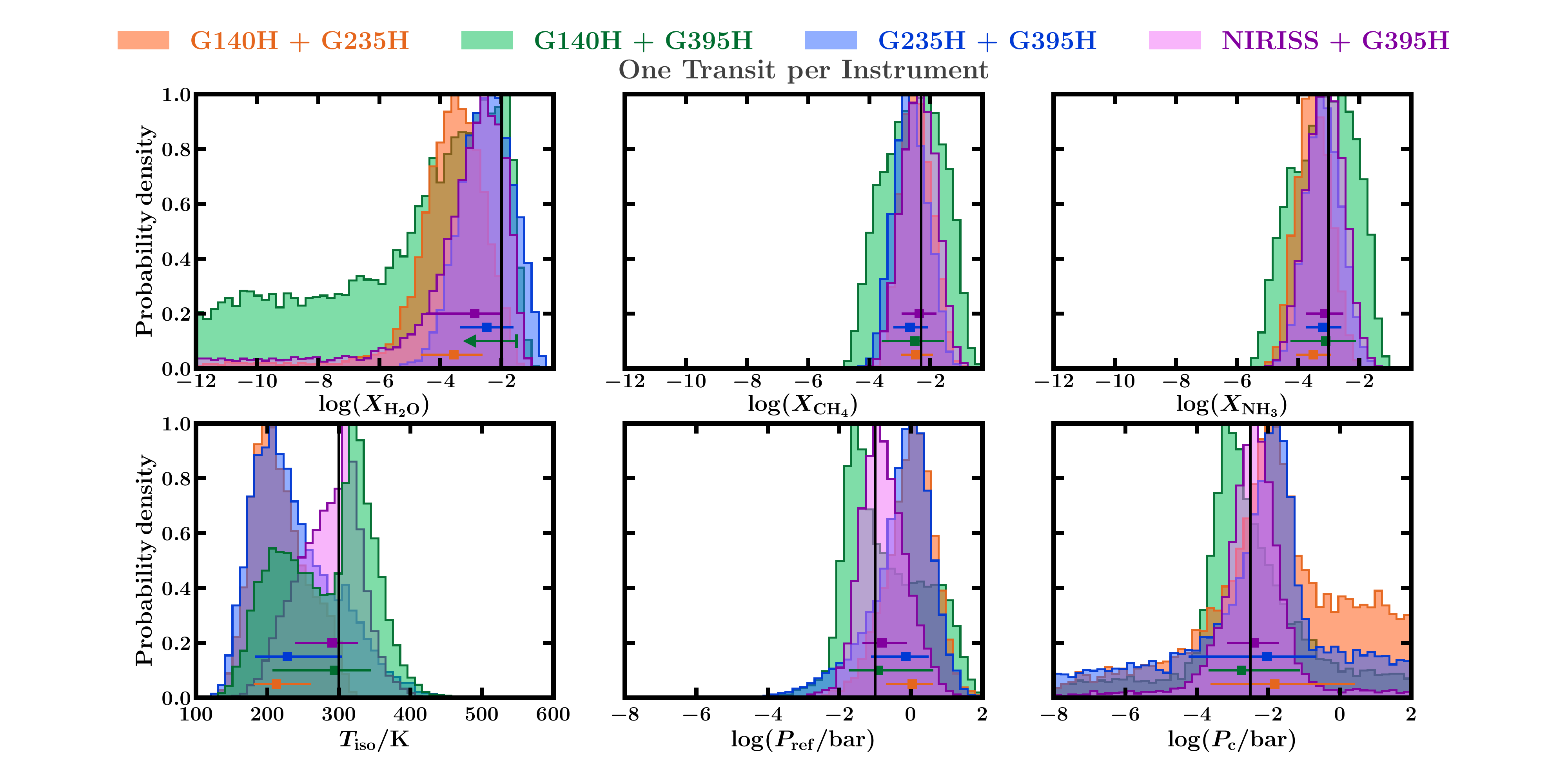}
    \caption{Posterior distributions for retrievals for the conservative case of K2-18~b with clouds, carried out on two-instrument combinations: NIRSpec G140H + G235H (orange), G140H + G395H (green), G235H + G395H (blue) as well as NIRISS + NIRSpec G395H. The input synthetic spectral data were generated for a nominal K2-18~b atmosphere at 10$\times$ solar elemental abundance with a cloud deck at 3~mbar. Black vertical lines denote the true values used to generate the data, while horizontal errorbars denote the median retrieved value and 1-$\sigma$ intervals. If a  retrieval results in a lower 1-$\sigma$ uncertainty larger than 2~dex, the errorbar is replaced with an arrow denoting the 2-$\sigma$ (95\%) confidence upper limit.}
    \label{fig:posterior_clouds_2transits}
\end{figure*}

As done previously, we run repeat retrievals to determine the variability of our results with different noise instances. We find that repeat retrievals on NIRSpec G140H data produce similar posterior distributions to those seen in figure \ref{fig:singleinstrument_clouds_posteriors}, where CH$_4$ and NH$_3$ mixing ratios are retrieved accurately while producing poorer H$_2$O constraints. Follow-up retrievals on NIRSpec G395H data were also unsuccessful at detecting H$_2$O, in some cases producing a posterior that is peaked at the true value but still has a substantial low abundance tail. The relative heights of the two peaks displayed by posteriors of CH$_4$, NH$_3$ and $T_\mathrm{iso}$ varied with different noise instances, with some retrievals obtaining only one of the two peaks.

Repeat retrievals on NIRSpec G235H and NIRISS also yield similar results to those shown in figure \ref{fig:singleinstrument_clouds_posteriors}. In a minority of cases, they succeed in fully constraining the abundances of H$_2$O and NH$_3$ in addition to CH$_4$, generally achieving precisions of $\sim$1~dex. This indicates that the success of retrievals on NIRISS and NIRSpec G235H observations in constraining the abundances of H$_2$O and NH$_3$ is dependent on specific spectral features not being degraded by random noise.

Overall, single-transit observations of K2-18~b consistently lead to CH$_4$ detections with all instrument configurations considered. Additionally, NH$_3$ is also detectable with some instruments, but its mixing ratio is not constrained as robustly. H$_2$O is the most challenging to retrieve for our particular model atmosphere, with the large majority of our retrievals only producing upper limits for its mixing ratio. We emphasise that these results are obtained with a minimal observing configuration, with only a single transit observed in each case. By observing more transits with each instrument or alternatively observing a planet that is more spectroscopically amenable than K2-18~b, better results can be obtained. Despite this, the mixing ratios of CH$_4$ and NH$_3$ can still be constrained when using NIRSpec G140H or G235H observations, although neither retrieval offers any firm indication that H$_2$O might also be present.

\subsubsection{Two-Instrument Observations}
\label{sec:two_transit_obs}

We now examine whether the addition of a second observation with a different, non-overlapping instrument is enough to allow retrievals to overcome the masking effects of clouds and robustly measure the abundances of H$_2$O, CH$_4$ and NH$_3$. In Figure \ref{fig:posterior_clouds_2transits}, we present the results of retrievals carried out on 4 instrument combinations: NIRSpec G140H + G235H, NIRSpec G140H + G395H, NIRspec G235H + G395H and NIRISS + NIRSpec G395H. We note that the NIRISS + NIRSpec G395H configuration achieves the same wavelength coverage as the three-instrument NIRSpec G140H + G235H + G395H combination considered in section \ref{sec:three_instrument_obs}, but at the cost of a lower precision and resolution over the $\sim$1-3~$\mu$m wavelength range covered by NIRISS.

As can be seen in figure \ref{fig:posterior_clouds_2transits}, all four retrievals perform better than those on single-transit observations, although ``tails'' towards low abundances can be seen in two of the H$_2$O posterior distributions. Retrieving on combined observations from NIRSpec G140H + G235H, which both individually led to constraints on the mixing ratios of CH$_4$ and NH$_3$ but not H$_2$O, again accurately constrained the mixing ratios of CH$_4$ and NH$_3$, this time with uncertainties of $\sim$0.5 and $\sim$0.6~dex, respectively. The most notable improvement is that the retrieval now also provides an estimate of the mixing ratio of H$_2$O with an uncertainty of $\sim$0.8~dex. Additionally, the retrieval is also more successful at constraining the cloud deck pressure, $P_\mathrm{c}$ compared to the individual retrievals on NIRSpec G140H or G235H observations, albeit with significant spread towards higher pressures that correspond to ``cloud free'' atmospheres.

The retrieval on NIRSpec G140H + G395H data produces tentative signs that H$_2$O may be present, with a peaked but unconstrained posterior distribution, which only offers an upper limit on the H$_2$O mixing ratio consistent with the true value. This retrieval does obtains accurate mixing ratio estimates for CH$_4$ and NH$_3$, but with the largest uncertainty of all two-instrument retrievals considered, at $\sim$~1.0 and $\sim$1.1~dex, respectively. Compared to the NIRSpec G140H + G235H retrieval, it obtains a more precise $P_\mathrm{c}$ estimate, but produces bimodal isotherm temperature and reference pressure posteriors, with one of the two modes corresponding to the true value. Notably, all other retrievals produce posteriors that have peaks corresponding to one of the two modes obtained by the NIRSpec G140H + G395H retrieval. This indicates that the retrievals are still susceptible to degeneracies between cloud truncation and scale height.

The retrieval on NIRSpec G235H + G395H data, obtains abundance constraints with a comparable precision to those obtained from NIRSpec G140H + G235H for all three molecules. The mixing ratio estimates obtained are $\mathrm{log}(X_{\mathrm{H}_{2} \mathrm{O}}) = -2.48^{+0.87}_{-0.89}$, $\mathrm{log}(X_{\mathrm{CH}_4}) = -2.67^{+0.59}_{-0.55}$ and $\mathrm{log}(X_{\mathrm{NH}_3}) = -3.18^{+0.60}_{-0.57}$. All are within 1-$\sigma$ of the respective true values.

Lastly, the retrieval on observations from NIRISS + NIRSpec G395H produces an estimate for the mixing ratio of H$_2$O, but with a slight tail in its posterior towards lower abundances. This leads to a lower 1-$\sigma$ interval to -1.6~dex, and a much smaller upper 1-$\sigma$ interval of +0.9~dex. Similarly to the other three retrievals, it also accurately retrieved the mixing ratios of CH$_4$ and NH$_3$, achieving precisions of $\sim$0.6~dex for both. Notably, it produces the most precise cloud deck constraint of all 2-instrument configurations considered, although the corresponding posterior distribution still shows a slight spread towards higher and lower pressures. This retrieval was also the most successful in constraining the isotherm temperature and reference pressure. This is due to the NIRISS + NIRSpec G395H configuration offering the largest wavelength coverage, encompassing the full $\sim$1-5$\mu$m range.

Across the 4 retrievals shown in figure \ref{fig:posterior_clouds_2transits}, the mixing ratios of CH$_4$ and NH$_3$ are consistently constrained, with all four retrievals producing estimates within 1-$\sigma$ of the true value. Two of the four retrievals produced H$_2$O mixing ratio posterior distributions with low abundance ``tails'', which indicate that the data do not entirely rule out models where the H$_2$O abundance is undetectably low. The NIRSpec G140H + G235H and G235H + G395H retrievals produces an H$_2$O posterior distribution without such a tail, corresponding to a precise abundance measurement. 

\begin{figure*}
    \centering
    \includegraphics[width=0.995\textwidth]{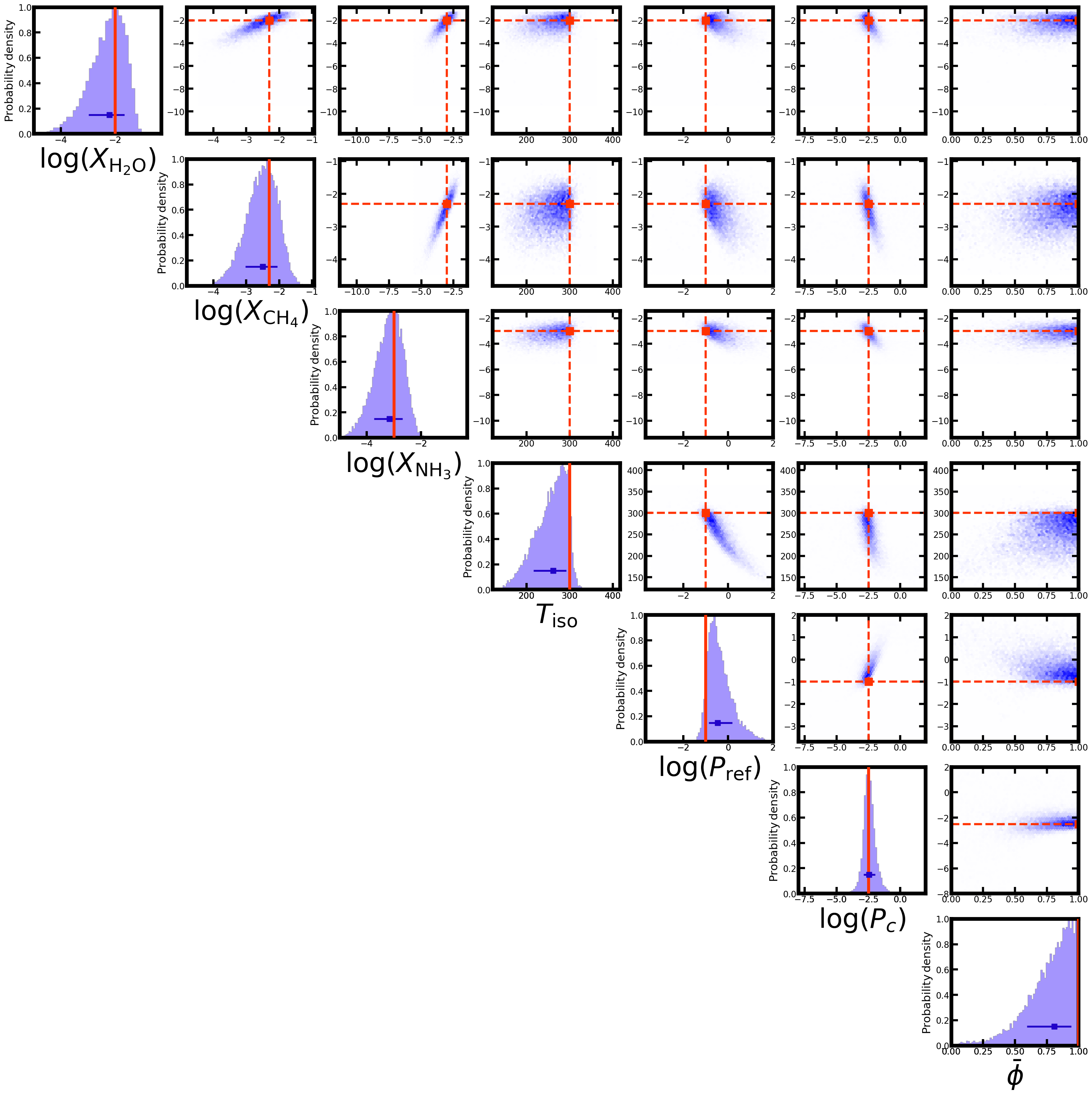}
    \caption{Posterior probability distribution for all 7 retrieved parameters obtained by a retrieval on combined NIRSpec G140H, G235H and G395H simulated observations of our cloudy K2-18~b case. The data were generated for an atmosphere which has a composition corresponding to 10$\times$ solar elemental abundances and a cloud deck present at 3~mbar, as described in section \ref{sec:canonical_model}. Horizontal errorbars denote the median and 1-$\sigma$ intervals, while red solid and dashed lines denote the true values used to generate the data.}
    \label{fig:threeinstrument_cloudy_corner}
\end{figure*}

As before we conduct repeat retrievals to assess the robustness of our results. Repeating the NIRSpec G140H + G235H retrieval produces similarly accurate CH$_4$ and NH$_3$ mixing ratio estimates, but sometimes also yield $\mathrm{log}(X_{\mathrm{H}_{2} \mathrm{O}})$ posteriors with low abundance tails. Repeat retrievals on observations with NIRSpec G140H + G395H obtain H$_2$O posteriors with low abundance tails of varying severity. Additionally, the relative heights of the bimodal isotherm temperature and reference pressure vary from run to run, with some exclusively finding one of the two peaks. Follow up retrievals on different noise instances of NIRISS and NIRSpec G395H data again produce H$_2$O posteriors with varying degrees of spread to lower mixing ratios, while reliably constraining the cloud deck pressure.

Repeat retrievals on re-generated observations with NIRSpec G235H + G395H also show some variability particularly for the retrieved H$_2$O abundance constraints. As can be seen in figure \ref{fig:contribution_plot}, there are two significant H$_2$O features that primarily drive the H$_2$O constraints from a NIRSpec G235H + G395H retrieval. Our findings therefore indicate that precision of the H$_2$O mixing ratio is conditional on these particular features not being too degraded by random noise. This risk can be mitigated by either observing a greater wavelength range, so as to include more molecular features, or repeat observations with the same instrument to improve the spectrophotometric precision achieved. This configuration however consistently retrieves the H$_2$O abundance more reliably and robustly than the NIRISS + NIRSpec G395H configuration. This indicates that in addition to wavelength coverage, the higher resolution and precision that NIRSpec G235H offers compared to NIRISS are also important in obtaining abundance estimates for cloudy atmospheres.

Given the significant challenge presented by our canonical cloudy model atmosphere of K2-18~b, with a cloud deck present at 3~mbar, we additionally consider cloud decks at lower altitudes. We find that with a cloud deck at 10~mbar all retrievals successfully constrain the mixing ratios of H$_2$O, CH$_4$ and NH$_3$ to within 1-$\sigma$ of the correct values. We therefore conclude that two-transit observations are viable for cloudy atmospheres where it is known a-priori that the cloud deck is at pressures greater than 3~mbar e.g. for K2-18~b \citep{Benneke2019, Madhusudhan2020}. We caution however that in the absence of such information, two-transit observations run the risk of failing to yield constraints on atmospheric properties, should a cloud deck at pressures less than 3~mbar (i.e., at higher altitudes) indeed be present.

Our findings therefore indicate that even with our highly challenging cloudy canonical model, two-instrument configurations, where each instrument is used to observe one transit, can successfully lead to constraints for the atmosphere's abundances and properties. Moreover, we find that this is possible with several two-instrument configurations, allowing observing programs to opt for whichever is optimal for their specific objectives. The NIRSpec G235H + G395H configurations however was the best performing, achieving the most precise abundance constraints and doing so more consistently than the other configurations. Our findings highlight the importance of wavelength coverage as well as spectroscopic precision in characterising cloudy atmospheres. In the following section, we examine to what extent retrieved atmospheric constraints are improved by increasing the number of instruments used to three, combining all three NIRSpec gratings.

\subsubsection{Three-Instrument Observations}
\label{sec:three_instrument_obs}

 \begin{figure*}
    \centering
    \includegraphics[width=\textwidth]{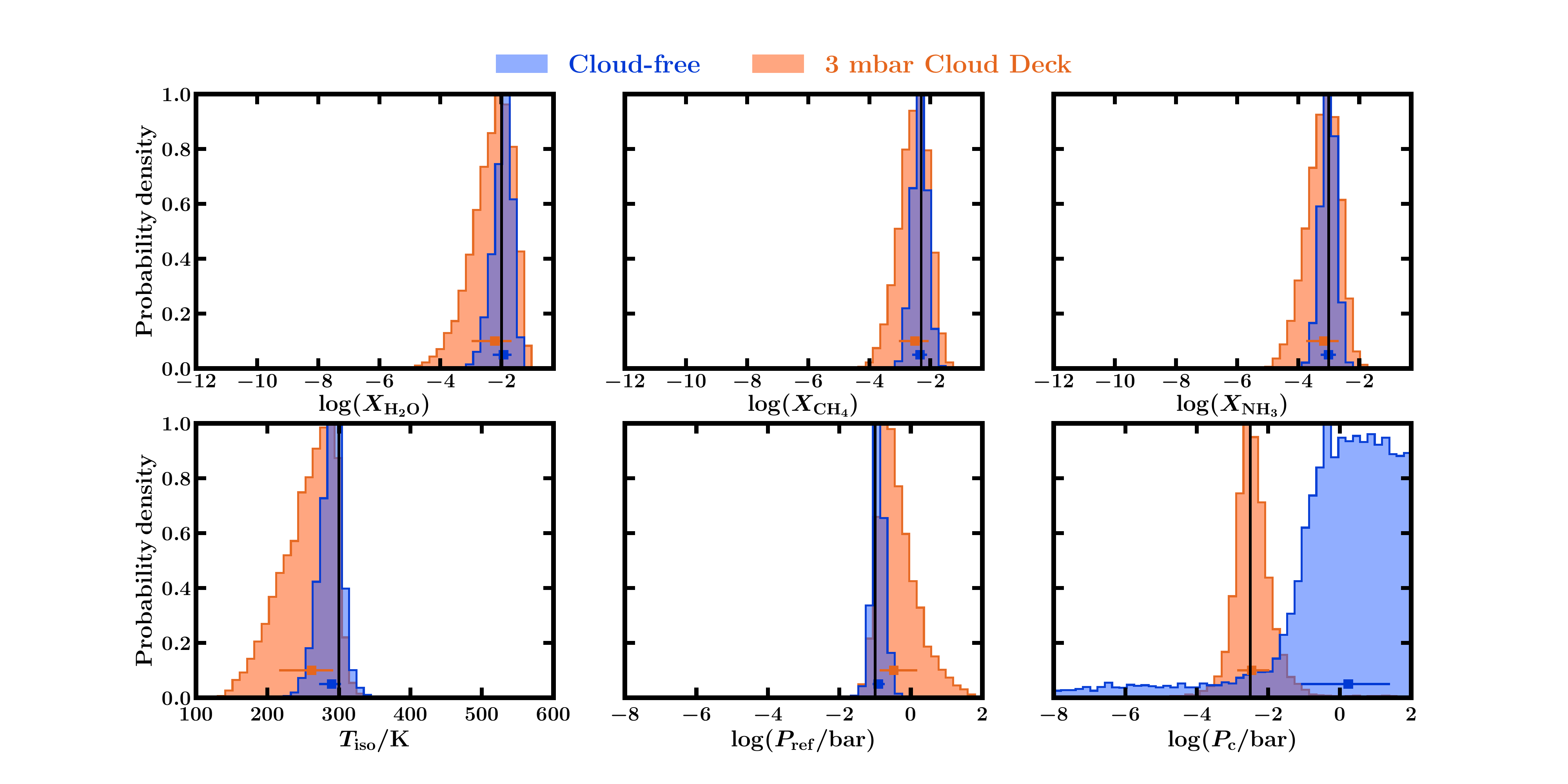}
    \caption{ Posterior distributions for retrievals for K2-18~b with no clouds (blue) and with a cloud deck present at 3~mbar (orange). Both retrievals were carried out on combined single-transit observations with three instrument settings: NIRSpec G140H, G235H and G395H. For both cases, the input simulated spectral data were generated for the canonical K2-18~b atmosphere at 10$\times$ solar elemental abundance with a cloud deck at 3~mbar as described in section \ref{sec:canonical_model}. Black vertical lines denote the true values used to generate the data, while horizontal errorbars denote the median retrieved value and 1-$\sigma$ intervals.}
    \label{fig:posteriors_cloudfree_vs_noclouds}
\end{figure*}

 We now consider the three-instrument configuration comprising of NIRSpec G140H + G235H + G395H,  which achieves a complete coverage of the $\sim$1-5~$\mu$m range, as shown in figure \ref{fig:contribution_plot}. We do not consider any other three-instrument combination, as NIRISS has a wavelength coverage that almost entirely overlaps with the NIRSpec G140H and G235H bands. We seek to understand the extent to which abundance constraints are improved in going from two- to three-instrument configurations. We therefore focus on two particular comparisons with results from section \ref{sec:two_transit_obs}. The first is with the NIRISS + NIRSpec G395H combination, which offers effectively the same $\sim$1-5~$\mu$m wavelength coverage as NIRSpec G140H + G235H + G395H, at the cost of NIRISS achieving a lower resolution and precision. The second is with NIRSpec G235H + G395H, which we found to be the best-performing two-instrument configuration in constraining atmospheric abundances, which does not cover the feature-right $\sim$1-2~$\mu$m range that NIRISS and NIRSpec G140H cover.
 
 The full posterior distribution obtained from our retrieval on the combined NIRSpec G140H + G235H + G395H observations is shown in figure \ref{fig:threeinstrument_cloudy_corner}. The retrieval successfully obtains accurate mixing ratio estimates, retrieving values of -2.22$^{+0.55}_{-0.77}$ for H$_2$O, -2.51$^{+0.45}_{-0.52}$ for CH$_4$ and -3.16$^{+0.48}_{-0.58}$ for NH$_3$. Additionally, the retrieval also successfully constrains the cloud deck pressure, with upper and lower 1-$\sigma$ intervals of +0.5 and -0.4~dex, respectively. The isotherm temperature and log-reference pressure are estimated to be 262$^{+32}_{-46}$~K and -0.48$^{+0.65}_{-0.41}$, both marginally more than 1-$\sigma$ away from the true value, but well within 2-$\sigma$.
 
 Retrieving on the three-instrument NIRSpec G140H + G235H + G395H configuration produces substantially more precise abundance estimates relative to any of the two-instrument configurations presented in section \ref{sec:two_transit_obs}, especially for H$_2$O. This is particularly notable when comparing to NIRSpec G235H + G395H, as the only difference is the inclusion of NIRSpec G140H, which on its own did not lead to any H$_2$O abundance constraints in section \ref{sec:single_transit_obs} or when combined with NIRSpec G395H in section \ref{sec:two_transit_obs}. The improvement in the H$_2$O mixing ratio precision that NIRSpec G140H provided indicates that it does contain information about the H$_2$O abundance, which can be accessed when combined with additional observations over a larger wavelength range. Additionally, the inclusion of NIRSpec G140H allows for more precise constraints for the mixing ratios of CH$_4$ and NH$_3$, which can indirectly lead to better H$_2$O constraints by alleviating degeneracies between the three molecules. The retrieval on the full three-instrument configuration also obtained more precise isotherm temperature and reference pressure estimates, than the NIRSpec G235H + G395H configuration. This supports earlier indications that observing a greater wavelength range and, hence, a greater number of absorption features, helps in establishing the spectrum's baseline in the absence of supporting optical data.
 
 Relative to the retrieval on NIRISS + NIRSpec G395H, we find that the improvements are mainly in the abundance estimates, due to the greater resolution and precision achieved by replacing NIRISS with NIRSpec G140H + G235H. Meanwhile, the isotherm temperature, reference pressure and cloud deck pressure are retrieved with comparable precision, again supporting the deduction that a wide wavelength range is beneficial in establishing the spectral baseline.

 Repeat retrievals with new noise instances show similar results, with comparable uncertainties to those shown in figure \ref{fig:threeinstrument_clear_corner}. In a minority of cases, the posterior distribution for the H$_2$O mixing ratio has a low abundance tail, indicating that H$_2$O is not well-retrieved. We find that H$_2$O being less precisely constrained tends to correlate with the cloud pressure and/or both the isotherm temperature and reference pressure being poorly retrieved as well. Moreover, we find that H$_2$O is well-retrieved more often than in the NIRSpec G235H +G395H case.
 
 We therefore find that there are meaningful gains to be had by observing with the NIRSpec G140H + G235H + G395H configuration over any two-instrument configuration. We note, however, that two-instrument configurations also succeed in constraining the abundances of all three dominant molecules, but less reliably and with a greater risk of the abundance estimates not being robust.
 
 We superpose the posterior distributions obtained for the cloudy and cloud-free cases in figure \ref{fig:posteriors_cloudfree_vs_noclouds}. It is evident that while the introduction of a cloud deck at 3~mbar has resulted in less precise constraints, all atmospheric parameters are retrieved well enough to allow for meaningful conclusions to be reached about the atmospheric properties of the planet. Moreover, observing additional transits, beyond the minimal single transit per instrument configuration explored in this work, as is the case for K2-18~b in Cycle 1 with programmes 2372 and 2722 (PI: Renyu Hu and Nikku Madhusudhan, respectively), will lead to even more precise constraints, approaching those obtained for the cloud-free case shown.

 Overall, our results indicate that even when a high-altitude cloud deck at 3~mbar is present in a temperate mini-Neptune like K2-18~b, orbiting a moderately bright M dwarf, a minimal JWST observing setup consisting of 1 transit per instrument and judiciously combining at least two instruments, can result in retrievals overcoming clouds and establishing highly precise abundance constraints. We note that our choice of cloud deck pressure in our canonical model is conservative, with the cloud top lying at lower pressures (i.e. higher altitudes) than the constraints obtained from HST observations and theoretical studies \citep{Benneke2019, Blain2021}. It is therefore likely that actual JWST observations of K2-18~b or similar planets with clouds will yield more precise constraints than those obtained in this work. As noted above, our choice of atmospheric composition consistently results in H$_2$O often having the poorest abundance constraints. The precision with which each molecule's mixing ratio is retrieved can be expected to vary depending on the specific atmospheric composition, which we explore in section \ref{sec:benchmark_depleted}.

\subsection{Case Study: TOI-732~c With Clouds}
\label{sec:case_study}

\begin{figure*}
    \includegraphics[width=\textwidth]{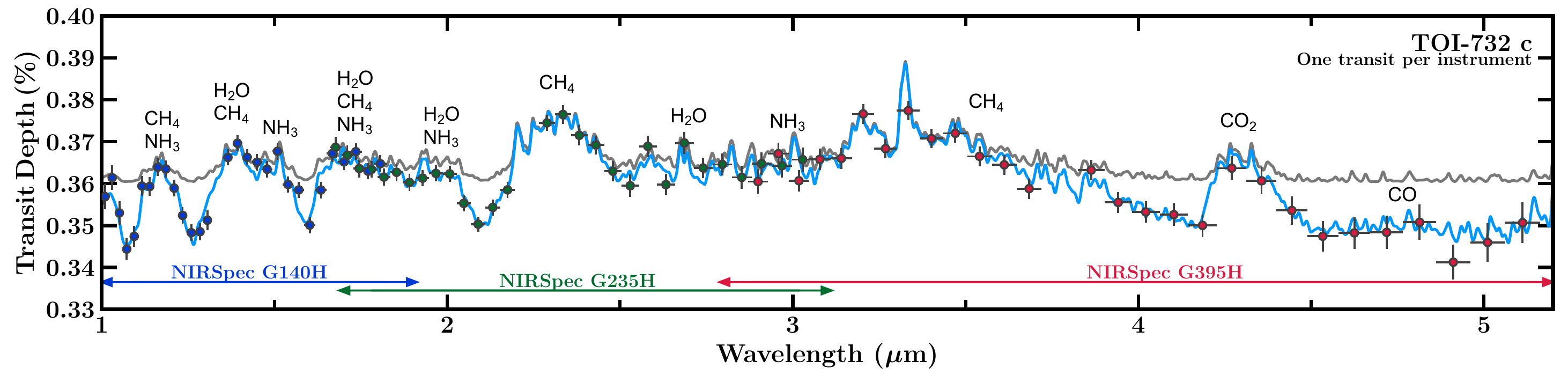}
    \caption{ Model transmission spectra of TOI-732~c a promising temperate mini-Neptune orbiting a bright M dwarf. The blue spectrum corresponds to the cloud free case of our canonical model described in section \ref{sec:canonical_model}, with the additional inclusion of CO and CO$_2$, both at 100 parts-per-million mixing ratio, for illustrative purposes. The grey spectrum is the same as the one in blue but with the addition of a 1~mbar cloud deck, which is one of the cloud deck pressures considered in section \ref{sec:benchmark_canonical}. The prominent spectral features are labelled with the corresponding molecules. Also shown are the corresponding JWST observations for the cloud-free model with NIRSpec G140H, G235H and G395H, each observing 1 transit, binned down to a resolution of R~=~50 for clarity.}
    \label{fig:TOI-732c_labelled_spectrum}
\end{figure*}

 In this section, we investigate the observing capabilities of JWST for a target more amenable to spectroscopic observations than K2-18~b. We focus on TOI-732~c \citep{Nowak2020, Cloutier2020}, a more recently-discovered temperate mini-Neptune, orbiting a star twice as bright as K2-18 in the $J$ band, as shown in table \ref{tab:planet_properties}. TOI-732~c has a somewhat higher equilibrium temperature compared to K2-18~b, at 363~K (for zero Bond albedo and full redistribution), as well as lower gravity. As a result, it has a larger atmospheric scale height compared to K2-18~b, and hence larger spectral features, which in addition to its brighter host star, leads to higher SNR observations. This can be seen in figure \ref{fig:TOI-732c_labelled_spectrum}, which shows our canonical model and corresponding simulated data for TOI-732~c, with each spectral feature labelled with the molecules giving rise to it. Motivated by the planet's higher equilibrium temperature, we use a nominal terminator temperature of 350~K, which is a 50~K increase from what was used for K2-18~b.

 We first examine how the improved SNR that TOI-732~c offers over K2-18~b affects the viability of single-transit single-instrument observations of a cloudy atmosphere in section \ref{sec:TOI-732_single_instruments}, using the same composition and 3~mbar cloud deck as used in section \ref{sec:wavelength_coverage}. In section \ref{sec:benchmark_canonical}, we use the instrument configuration that led to the best abundance constraints for K2-18~b, specifically NIRSpec G140H + G235H + G395H, to examine how the detectability of H$_2$O, CH$_4$ and NH$_3$ varies with cloud deck pressure. We now also include cases where the cloud deck is at lower pressures (i.e. higher altitudes) than 3~mbar, the value used so far in this work. We then carry out the same investigation in section \ref{sec:benchmark_depleted} for an atmospheric composition where CH$_4$ and NH$_3$ are depleted by 1~dex relative to the canonical model. In doing so, we explore how atmospheric composition affects the retrieved abundance constraints of the three dominant molecules considered, as well as how severely spectral features are masked by clouds. All retrieved mixing ratio constraints for TOI-732~c are summarised in table \ref{tab:TOI-732c_results}.

\begin{table}
    \centering
    \caption{Retrieved log-mixing ratio constraints for TOI-732~c for all atmospheric and instrumental configurations considered.}
    \begin{tabular}{l|c|c|c}
        & \multicolumn{3}{c}{log-Mixing Ratios} \\[0.5mm]
        Case &  H$_2$O & CH$_4$ & NH$_3$  \\[0.5mm]
        \hline
        \hline
        \multicolumn{4}{c}{Canonical Abundances} \\[0.5mm]
        \hline
        True Values& -2 & -2.3 & -3 \\[0.5mm]
        \hline
        \multicolumn{4}{l}{\emph{Single-Instrument Configurations}} \\[0.5mm]
        \emph{3~mbar Cloud Deck} \\[0.5mm]
        NIRSpec G140H                  &       $(-1.89)$                 &   $-2.39^{+0.97}_{-0.93}$    &    $-3.41^{+0.98}_{-0.96}$ \\[0.5mm]
        NIRSpec G235H                   &   $-2.45^{+0.71}_{-0.60}$     &   $2.62^{+0.49}_{-0.41}$     &    $-3.33^{+0.56}_{-0.49}$ \\[0.5mm]
        NIRSpec G395H                   &       $(-0.88)$                 &   $-1.50^{+0.49}_{-1.23}$    &    $-2.85^{+0.78}_{-1.32}$  \\[0.5mm]
        NIRISS                          &   $-1.93^{+0.65}_{-0.70}$     &   $-2.48^{+0.50}_{-0.47}$    &    $-3.46^{+0.51}_{-0.50}$ \\[0.5mm]
        \\[0.5mm]
        \multicolumn{4}{l}{\emph{Three-Instrument Configuration}} \\[0.5mm]
        Cloud-Free                     &   $-1.95^{+0.29}_{-0.40}$     &   $2.26^{+0.20}_{-0.25}$    &      $-2.89^{+0.20}_{-0.26}$\\[0.5mm]
        10$^{-3}$~bar Cloud Deck       &   $-1.69^{+0.37}_{-0.87}$     &   $-2.35^{+0.31}_{-0.60}$    &     $-3.45^{+0.39}_{-0.66}$ \\[0.5mm]
        10$^{-4}$~bar Cloud Deck       &   $-2.63^{+0.64}_{-0.79}$     &   $-2.91^{+0.55}_{-0.65}$    &     $-3.82^{+0.68}_{-0.87}$  \\[0.5mm]
        10$^{-4.5}$~bar Cloud Deck      &   $(-2.62)$     &    $-2.76^{+0.65}_{-0.86}$   &   $(-3.37)$  \\[0.5mm]
        
        \hline
        \hline
        \multicolumn{4}{c}{Depleted CH$_4$ and NH$_3$} \\[0.5mm]
        \hline
        True Values& -2 & -3.3 & -4 \\[0.5mm]
        \hline
        \multicolumn{4}{l}{\emph{Three-Instrument Configuration}} \\[0.5mm]
        Cloud-Free                      &   $-2.07^{+0.28}_{-0.28}$     &   $-3.34^{+0.19}_{-0.19}$     &   $-4.01^{+0.19}_{-0.20}$     \\[0.5mm]
        10$^{-2}$~bar Cloud Deck        &   $-2.11^{+0.62}_{-0.59}$     &   $-3.21^{+0.40}_{-0.39}$     &   $-4.31^{+0.40}_{-0.41}$  \\[0.5mm]
        10$^{-3}$~bar Cloud Deck        &   $-2.42^{+0.76}_{-0.80}$     &   $-3.24^{+0.59}_{-0.58}$     &   $-3.83^{+0.60}_{-0.60}$  \\[0.5mm]
        10$^{-4}$~bar Cloud Deck        &   $-2.97^{+0.70}_{-1.24}$     &   $-4.16^{+0.71}_{-1.10}$     &   $(-3.92)$  \\[0.5mm]
        \hline

    \end{tabular}
    \newline
    \footnotesize{Note: In cases where the lower 1-$\sigma$ interval spans more than 2 dex, we instead list the 2-$\sigma$ (95\%) upper estimate in brackets.}
    \label{tab:TOI-732c_results}
\end{table}

\subsubsection{Single-Instrument Observations}
\label{sec:TOI-732_single_instruments}

\begin{figure*}
    \centering
    \includegraphics[width=\textwidth]{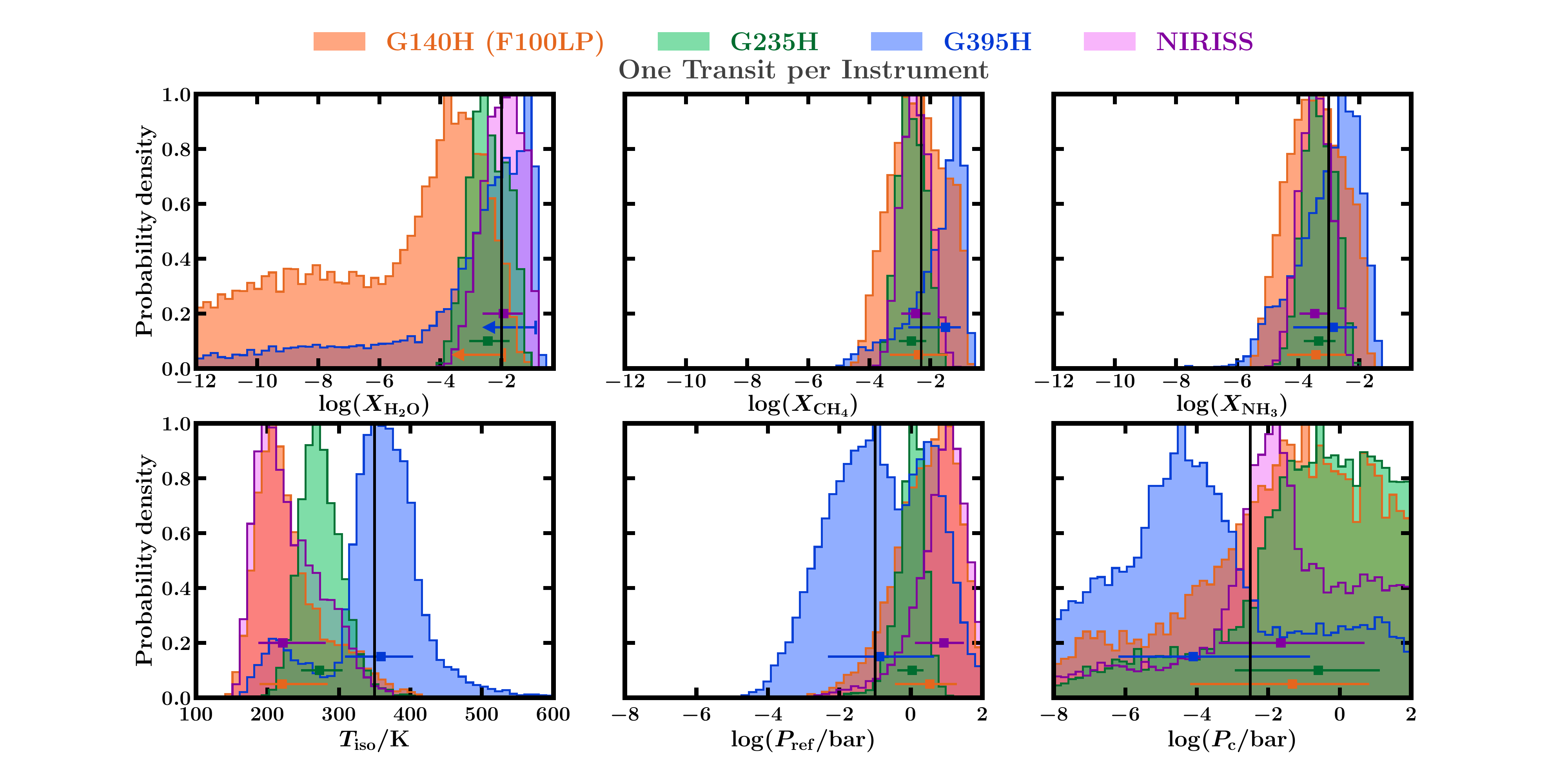}
    \caption{ Posterior distributions of retrievals carried out for the more optimistic case of TOI-732~c with our canonical cloudy model atmosphere, retrieving on simulated single-instrument observations with NIRSpec G140H (orange), G235H (green) and G395H (blue) as well as NIRISS (purple), retrieving on data from one instrument at a time. All data were generated for a nominal atmosphere at 10$\times$ solar elemental abundance with a cloud deck at 3~mbar. Black vertical lines denote the true values used to generate the data, while horizontal errorbars denote the median value and 1-$\sigma$ intervals. If a retrieval produces a lower 1-$\sigma$ interval spanning more than 2~dex, the errorbar is replaced with an arrow denoting the 2-$\sigma$ (95\%) confidence upper limit. }
    \label{fig:TOI-732c_singleinstruments}
\end{figure*}

In this section, we explore how retrievals on the same single-transit single-instrument configurations perform for the more representative case of TOI-732~c. Figure \ref{fig:TOI-732c_singleinstruments} shows the retrieved posterior distributions for all four instrument configurations. Comparing these results to the equivalent for K2-18~b in figure \ref{fig:singleinstrument_clouds_posteriors}, we find that there is significant improvement in the abundance constraints the retrievals produce, particularly for H$_2$O and NH$_3$.

Both NIRSpec G235H and NIRISS observations lead to precise constraints for all three molecules. With NIRSpec G235H, the retrieved log-mixing ratio estimates for H$_2$O, CH$_4$ and NH$_3$ are, $-2.45^{+0.71}_{-0.60}$, $-2.62^{+0.49}_{-0.41}$ and $-3.33^{+0.56}_{-0.49}$, respectively. Using NIRISS data, the log-mixing ratio estimates obtained of H$_2$O, CH$_4$ and NH$_3$ are $-1.93^{+0.65}_{-0.70}$, $-2.48^{+0.50}_{-0.47}$ and $-3.46^{+0.51}_{-0.50}$, respectively, which are marginally less precise than those obtained using NIRSpec G235H. NIRSpec G140H and NIRSpec G395H also lead to CH$_4$ and NH$_3$ detections. With NIRSpec G140H data, our retrievals constrained CH$_4$ and NH$_3$ with an uncertainty of $\sim$1~dex for both molecules. The retrieval on NIRspec G395H data produced upper and lower 1-$\sigma$ intervals of +0.5 and -1.2~dex, respectively, for CH$_4$ and +0.8 and -1.3~dex, respectively, for NH$_3$. While neither retrieval was as successful in constraining the mixing-ratio of H$_2$O as the NIRISS and NIRSpec G235H retrievals, the NIRSpec G140H and G395H posterior distributions obtained are more strongly peaked close to the correct value than the equivalent posteriors obtained for K2-18~b.

The better performance of the retrievals shown in this section compared to those in section \ref{sec:single_transit_obs} is driven by a substantial improvement in SNR, thanks to TOI-732~c having a brighter host star than K2-18~b, as well as having larger spectral features. We highlight that a similar improvement can also be achieved for a K2-18~b-like planet by observing more than 1 transit with a given instrument. Repeat retrievals on new data instances for all three NIRSpec gratings produce similar results to those shown in figure \ref{fig:TOI-732c_singleinstruments}. Retrievals on NIRISS data show some variability, in some cases obtaining H$_2$O posteriors with spread towards lower abundances.

\subsubsection{Three Instruments}
\label{sec:benchmark_canonical}

\begin{figure*}
    \centering
    \includegraphics[width=\textwidth]{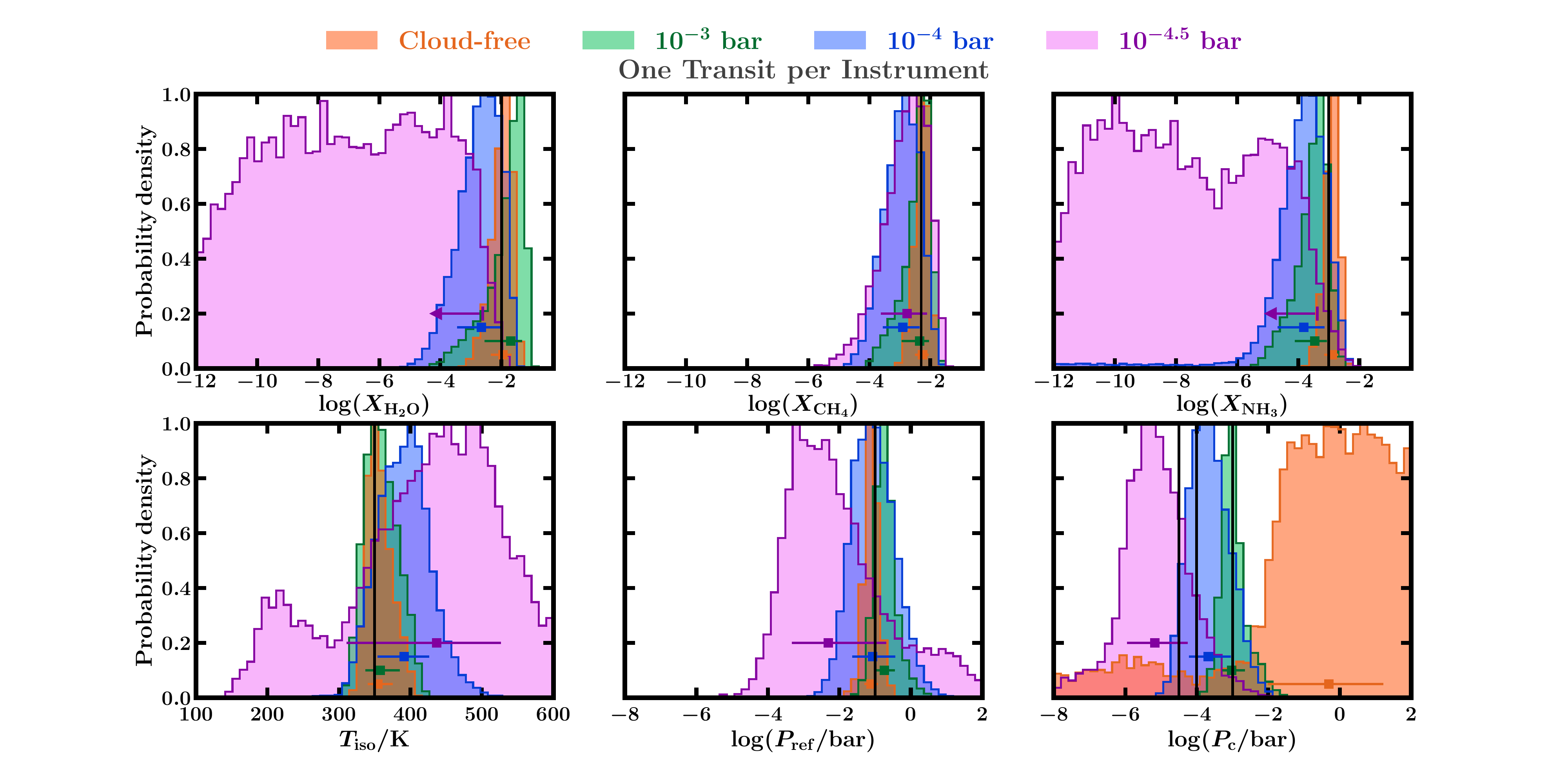}
    \caption{Posterior distributions for retrievals carried out on synthetic data for TOI-732~c, a more spectroscopically favourable planet than K2-18~b, with H$_2$O, CH$_4$ and NH$_3$ at 10$\times$ solar abundances and cloud decks at pressures of 10$^{-3}$~bar (green),  10$^{-4}$~bar (blue) and 10$^{-4.5}$~bar (purple), as well as without a cloud deck present (orange). All four retrievals were carried out on combined simulated observations from NIRSpec G140H, G235H and G395H.  Black vertical lines denote the true values used to generate the data, while errorbars denote each parameter's median retrieved values and corresponding 1-$\sigma$ intervals obtained in each of the retrievals. In cases where the retrieval only finds an upper limit for a molecular abundance, the errorbar is replaced with a arrow denoting the 2-$\sigma$ confidence upper limit.}
    \label{fig:TOI-732c_benchmark_equilibrium}
\end{figure*}

We now consider retrievals on TOI-732~c observations with the three-instrument NIRSpec G140H + G235H + G395H configuration for different cloud deck pressures. The posterior distributions for four retrievals, carried out for different cloud deck pressures up to $10^{-4.5}$~bar (0.03~mbar), are presented in figure \ref{fig:TOI-732c_benchmark_equilibrium}. Starting with the base case of a cloud-free atmosphere, we find that our retrieval obtains atmospheric parameter constraints with better precision than those obtained for K2-18~b in section \ref{sec:cloud_free_reference}. Specifically, we retrieve log-mixing ratio estimates of -1.95$^{+0.29}_{-0.40}$ for H$_2$O, -2.26$^{+0.20}_{-0.25}$ for CH$_4$ and -2.89$^{+0.20}_{-0.26}$ for NH$_3$. For the case where the cloud deck is at 10$^{-3}$~bar (1~mbar), an even lower pressure than the 3~mbar considered in section \ref{sec:wavelength_coverage}, we once again obtain good constraints for the mixing ratios of all three molecules, with upper and lower 1-$\sigma$ intervals of +0.4 and -0.9 for H$_2$O, +0.3 and -0.6 for CH$_4$ and +0.4 and -0.7 for NH$_3$. These uncertainties are comparable to those obtained in section \ref{sec:three_instrument_obs} for K2-18~b with a less obstructive cloud deck at 3~mbar, thanks to the significantly improved SNR that TOI-732~c offers. 

We find that reliable abundance constraints for all three molecules in TOI-732~c can be obtained for cloud deck pressures as low as 10$^{-4}$~bar (0.1~mbar). As shown in blue in figure \ref{fig:TOI-732c_benchmark_equilibrium}, we find that the mixing ratios of all three molecules are well retrieved, albeit at a lower precision than in the 1~mbar case. Our retrieval obtains upper and lower 1-$\sigma$ interval values of +0.6 and -0.8~dex for H$_2$O, +0.6 and -0.7~dex for CH$_4$ and +0.7 and -0.9~dex for NH$_3$. Additionally, NH$_3$ shows a slight low abundance tail. 
Having a cloud deck at $10^{-4.5}$~bar (0.03~mbar), shown in purple in figure \ref{fig:TOI-732c_benchmark_equilibrium}, precludes good  constraints on H$_2$O and NH$_3$. Nevertheless, the CH$_4$ mixing ratio is still retrieved with uncertainties that are only slightly larger than those obtained in the 0.1~mbar case. We note that the retrieved isotherm temperature is now showing two broad peaks, one at $\sim$200~K and one at $\sim$500~K. This indicates that the retrieval is being led astray by degeneracies, which is why it fails to constrain both H$_2$O and NH$_3$, which are both less spectrally dominant than CH$_4$. We additionally considered the case of a cloud-deck at 10$^{-5}$~bar and found that none of the three mixing ratios can be meaningfully constrained.

We highlight that these abundance constraints in all cases were obtained by retrieving on unbinned native-resolution (R$\sim$2700) NIRSpec data. By avoiding binning, retrievals can take advantage of strong absorption features from H$_2$O, CH$_4$ and NH$_3$ that remain observable at the native resolution despite the significant truncation from the cloud deck. This is particularly the case for $P_c = $ 0.1 mbar which provides good constraints on all the three molecules despite the presence of a high altitude cloud-deck. Our findings are the medium-resolution equivalent of prior reports in literature that high-resolution (R$\gtrsim$25,000) observations are capable of overcoming clouds in mini-Neptune atmospheres \citep{Gandhi2020, Hood2020}.

We also explore the impact of binning the observed spectra on the retrieved abundance constraints for cloudy atmospheres. We find that for cases with very high altitude clouds, excessive binning of spectra could lead to information loss resulting in less precise and less accurate abundance estimates. For example, considering the $P_\mathrm{c} = $ 0.1 mbar case of TOI-732~c, we find that binning to R=50 results in a largely unconstrained H$_2$O abundance and an NH$_3$ posterior with a significant low abundance tail. On the other hand, binning to R$\gtrsim$500 allows meaningful constraints on all three molecules approaching those achieved with native resolution spectra. The effect of binning is reduced for atmospheres that are cloud-free or with lower-altitude clouds.

\begin{figure*}
    \centering
    \includegraphics[width=\textwidth]{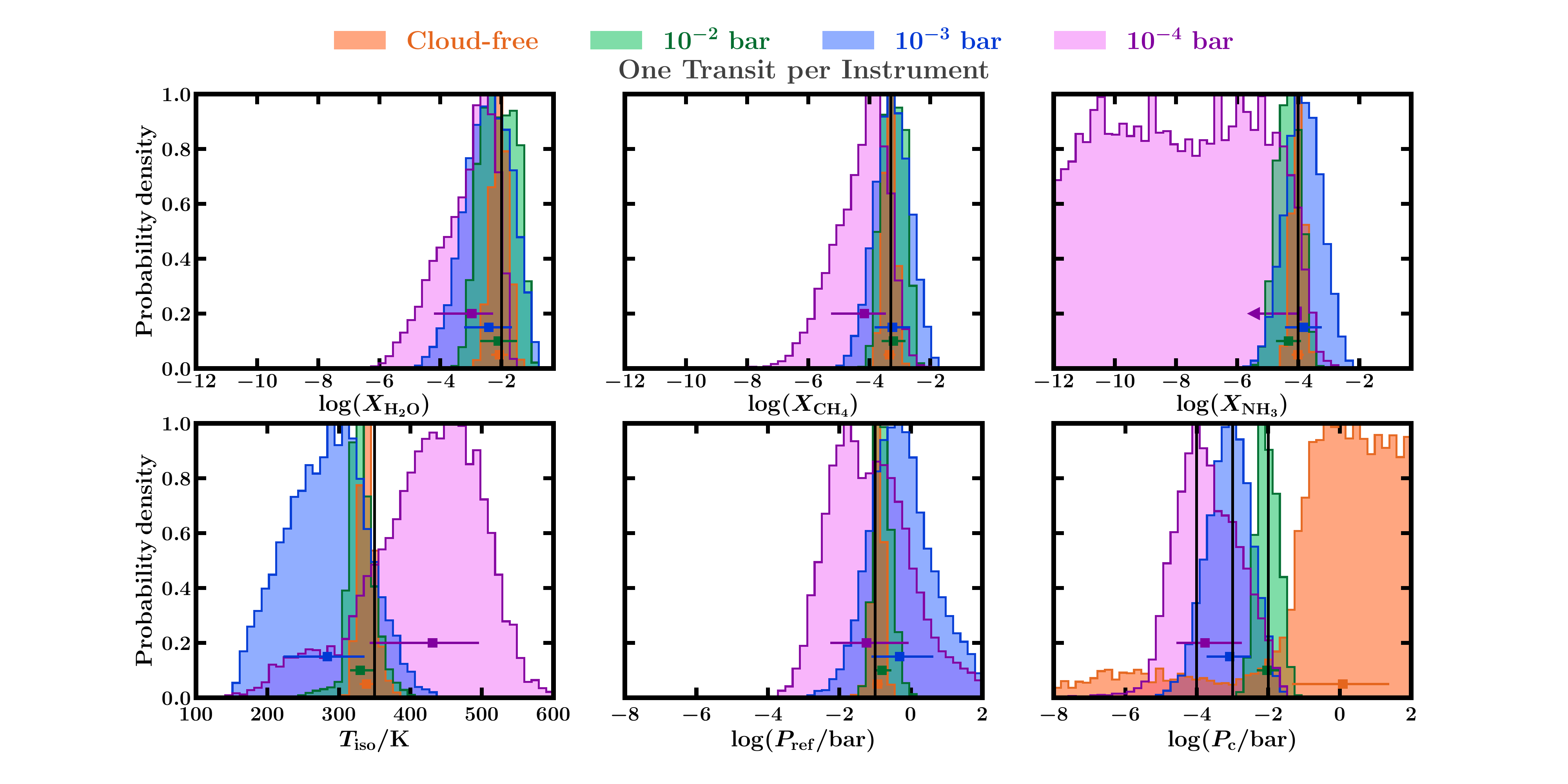}
    \caption{ Posterior distributions for retrievals of TOI-732~c with depleted CH$_4$ and NH$_3$ abundances relative to the canonical model. The volume mixing ratios of H$_2$O, CH$_4$ and NH$_3$ are $10^{-2}$, 5$\times 10^{-4}$ and $10^{-4}$, respectively, the latter two being lower by 1~dex compared to the canonical model used in rest of this work. Each retrieval corresponds to a different cloud deck pressure: 10$^{-3}$~bar (green), 10$^{-4}$~bar (blue), 10$^{-4.5}$~bar (purple) and a cloud-free case (orange). All four retrievals were carried out on combined simulated observations with NIRSpec G140H, G235H and G395H. Black vertical lines denote the true values used to generate the data, while points with error bars denote each parameter's median retrieved values and corresponding 1-$\sigma$ intervals. In cases where the retrieval only finds an upper limit for a molecular abundance, the error bar is replaced with an arrow denoting the 2-$\sigma$ upper limit. }
    \label{fig:TOI-732c_benchmark_depletion}
\end{figure*}

We carry out repeat retrievals on different noise instances to ensure that our results are not the result of noise-specific features. We confirm that repeat retrievals carried out with no clouds present are consistently successful at retrieving all atmospheric parameters, with the expected statistical variability, i.e. the majority of retrieved estimates lying within 1-$\sigma$ of true values while a minority between 1- and 2-$\sigma$. The same holds true for the 10$^{-3}$~bar case. Repeating retrievals with a cloud deck at 10$^{-4}$~bar, we find that in some cases H$_2$O and NH$_3$ display low abundance ``tails'' of varying severity, indicating that both are at the limit of observability. In a small minority of noise instances, H$_2$O is wholly unconstrained. Additionally, repeat retrievals with a $10^{-4.5}$~bar cloud deck sometimes find slight indications that H$_2$O or NH$_3$ may be present in the form of posterior distributions that are largely unconstrained but notably peaked near the true value. Encouragingly, the cloud deck pressure is consistently retrieved successfully, which means that it is possible to diagnose that the H$_2$O and NH$_3$ non-detections are due to high altitude clouds, and the solution to detecting them would be to observe additional transits to improve the spectroscopic precision of the data.

We therefore find that native-resolution JWST observations of mini-Neptunes more amenable to spectroscopic observation than K2-18~b, such as TOI-732~c, can lead to atmospheric abundance constraints even in the presence of clouds at pressures lower than the 3~mbar considered in section \ref{sec:wavelength_coverage}. The minimum cloud deck pressure that each molecule is detectable at, however, varies. For the atmospheric composition considered here, the mixing ratio of H$_2$O and NH$_3$ can no longer be constrained for cloud deck pressures of $10^{-4.5}$ (0.03~mbar) and below, while our retrievals can still estimate the abundance of CH$_4$ even with a cloud deck at the same cloud deck pressure. The minimum cloud deck pressure that each molecule is detectable is also expected to vary depending on the specific atmospheric composition. Exploring this is the focus of the next section.

\subsubsection{Three Instruments + Depleted CH$_4$ and NH$_3$}
\label{sec:benchmark_depleted}

  We now carry out the same benchmark as before, this time deviating from the canonical abundances used so far in this work. Specifically, we consider a new atmospheric composition, with CH$_4$ and NH$_3$ at mixing ratios of 5$\times$10$^{-4}$ and 10$^{-4}$, respectively, i.e. depleted by 1~dex relative to the canonical model, and the same 0.01 H$_2$O mixing ratio as before. As discussed in section \ref{sec:canonical_model}, this scenario corresponds to an atmosphere where atmospheric processes have either enhanced the abundance of H$_2$O or reduced the abundances of CH$_4$ and NH$_3$.

  As seen in figure \ref{fig:contribution_plot}, which corresponds to the canonical 10$\times$~solar composition, CH$_4$ and NH$_3$ have significant spectral contributions which often mask those from H$_2$O. As a result, the H$_2$O mixing ratio is consistently the least precisely constrained in sections \ref{sec:TOI-732_single_instruments} and \ref{fig:TOI-732c_benchmark_equilibrium}. Now that CH$_4$ and NH$_3$ are less abundant, it can therefore be expected that the H$_2$O mixing ratio will be more precisely retrieved. Conversely, CH$_4$ and especially NH$_3$, which has weaker spectral features than CH$_4$, are now expected to be the ones that are less easily constrained.
 
  Figure \ref{fig:TOI-732c_benchmark_depletion} shows the obtained posterior distributions for the cloud free case as well as when clouds are present at pressures of 10$^{-2}$, 10$^{-3}$ and 10$^{-4}$~bar. In the cloud-free case, the retrieved log-mixing ratios are -2.07$^{+0.28}_{-0.28}$ for H$_2$O, -3.34$^{+0.19}_{-0.19}$ for CH$_4$ and -4.01$^{+0.19}_{-0.20}$ for NH$_3$. Notably, all three are retrieved more precisely than in the non-depleted case presented in section \ref{sec:benchmark_canonical}. This indicates that H$_2$O features being more prominent not only help in obtaining more precise H$_2$O abundances, but also help improve the precision of the CH$_4$ and NH$_3$ estimates. This is due to the inherent degeneracy between molecular abundances when absorption features partially mask each other, which is alleviated when CH$_4$ and NH$_3$ are at lower abundances.
  
  Both the 10$^{-2}$ and 10$^{-3}$~bar cloud deck retrievals succeeded in constraining the mixing ratios of all three molecules. With a cloud deck present at 10$^{-2}$~bar, the H$_2$O mixing ratio is retrieved with a precision of 0.6~dex while CH$_4$ and NH$_3$ are both retrieved with 0.4~dex precision. This is a significant deterioration compared to the constraints obtained in the cloud-free case. Setting the cloud deck pressure to 10$^{-3}$~bar, the retrieval again obtain accurate mixing ratio estimates, with an uncertainty 0.8~dex for H$_2$O and 0.6~dex for CH$_4$ and NH$_3$. These uncertainties are greater than for the equivalent 10$^{-3}$~bar cloud deck retrieval for the canonical abundance case, indicative of the greater impact that a cloud deck can have on a transmission spectrum at lower atmospheric abundances.
  
  Lastly, when the cloud deck lies at 10$^{-4}$~bar, the retrieval is no longer able to constrain the mixing ratio of NH$_3$. The retrieval is able to offer constraints for the H$_2$O and CH$_4$ mixing ratios, with both lying more than 1-$\sigma$ away from the true values but are still within 2-$\sigma$. Specifically, H$_2$O is retrieved with upper and lower 1-$\sigma$ intervals of +0.7 and -1.2~dex, respectively, while for CH$_4$ the upper and lower 1-$\sigma$ intervals are +0.7 and -1.1.
  
  Similarly to before, we carry out repeat retrievals on new noise instances to establish the repeatability of our findings. Both the cloud-free and 10$^{-2}$~bar cloud deck cases produce consistent results, with all three molecules retrieved with precisions comparable to those described above, and the majority of estimates lying within 1-$\sigma$ of the true value and a minority lying between 1- and 2-$\sigma$. For the 10$^{-3}$~bar cloud deck case, retrievals occasionally produce a low abundance tail of varying severity in the NH$_3$ posterior distribution. This indicates that in order to constrain the abundance of NH$_3$, there must be no significant degradation of NH$_3$ spectral features by random noise. Lastly, repeat retrievals with a 10$^{-4}$~bar cloud deck are sometimes able to produce partially constrained NH$_3$ posteriors that are peaked close to the true values, offering a tentative indication that NH$_3$ may be present. In several cases,we obtain H$_2$O and CH$_4$ posteriors that also display low abundance ``tails'', indicating that both are at the limit of observability.
  
  Our results therefore show that abundance constraints for all three dominant molecules are still achievable when CH$_4$ and NH$_3$ are depleted relative to H$_2$O, for cloud top pressures down to $\sim 10^{-3}$~bar. We find that in the non-cloudy case, depletion of CH$_4$ and NH$_3$ leads to even more precise abundance constraints not just for H$_2$O but also for CH$_4$ and NH$_3$. The precision to which these abundance measurements are made however is more sensitive to cloud deck pressure, showing more rapid deterioration as clouds are placed at higher and higher altitudes. We find that NH$_3$ remains detectable down to a cloud deck pressure of 10$^{-3}$~bar, while H$_2$O and CH$_4$ can be detected in atmospheres with 10$^{-4}$~bar cloud decks. Extrapolating from our findings, it can be expected that for mini-Neptune atmospheres that are more enriched than those used in this work, high-altitude clouds will have a comparatively lesser impact on the transmission spectrum.

\section{Summary and Discussion}
\label{sec:conclusion}

 The atmospheric characterisation of temperate, low-mass exoplanets is a major frontier of the JWST era. In this work, we investigate the potential of transit spectroscopy with JWST for characterising the atmospheric compositions of temperate, cloudy mini-Neptunes. We first examine how precisely atmospheric abundances can be retrieved in the cloud-free case, and how these constraints are affected when high-altitude clouds are present. Additionally, we investigate what abundance constraints are achievable with different JWST instrument combinations. We also explore the constraints possible for different atmospheric compositions, highlighting the interplay between key molecular features. We pursue these investigations for two mini-Neptune prototypes, K2-18~b and TOI-732~c, simulating observations with JWST NIRISS and the three high-resolution NIRSpec gratings G140H, G235H and G395H. In all the cases, we assume a canonical atmospheric composition with 10$\times$~solar metallicity and single-transit observations with each instrument considered. 
 
 Our primary finding is that JWST transmission spectroscopy of temperate mini-Neptunes orbiting bright M dwarfs can provide precise constraints on their atmospheric compositions with modest observing time even in the presence of clouds at significantly high altitudes. In what follows, we summarise our main results, beginning with K2-18~b, a habitable-zone mini-Neptune orbiting an M2.5 dwarf, chosen as a relatively conservative case.

\begin{itemize}

\item For a cloud-free K2-18~b, single-transit, single-instrument observations with NIRISS or NIRSpec G235H can provide abundance constraints for H$_2$O, CH$_4$ and NH$_3$ with precisions better than $\sim$0.8~dex. A three-instrument combination using NIRspec G140H + G235H + G395H, in the 1-5 $\mu$m range provides constraints better than $\sim$0.3~dex for all three molecules.

\item For a cloudy K2-18~b multiple observations are required to obtain precise and robust abundance constraints. Single-transit three-instrument observations using NIRSpec G140H + G235H + G395H lead to precise abundance constraints for cloud top pressures as low as 3~mbar. H$_2$O, CH$_4$ and NH$_3$ are retrieved with precisions of $\sim$0.7, $\sim$0.5 and $\sim$0.6~dex, respectively.

\item Considering two-instrument combinations, NIRSpec G235H + G395H provides the best  constraints, with precisions of $\sim$0.9~dex for H$_2$O and $\sim$0.6~dex for both CH$_4$ and NH$_3$, assuming the same cloud-top pressure as above. 

\end{itemize}

We then consider a cloudy TOI-732~c, a more observationally favourable mini-Neptune due to its higher temperature, lower gravity, and smaller and brighter host star compared to K2-18~b.

\begin{itemize}

    \item With a cloud deck at 3~mbar, single-instrument observations with NIRISS or NIRSpec G235H lead to good constraints on all three molecules, with precisions $\lesssim$0.7~dex for H$_2$O and $\lesssim$0.5~dex for CH$_4$ and NH$_3$.
    
    \item Furthermore, using a three-instrument combination with NIRSpec G140H + G235H + G395H, H$_2$O and NH$_3$ can be constrained precisely with cloud decks at pressures as low as 0.1~mbar, while CH$_4$ constraints remain possible for cloud deck pressures down to 0.03~mbar.
    
    \item Depleting CH$_4$ and NH$_3$ by 1~dex relative to the canonical 10$\times$~solar case, the mixing ratio of NH$_3$ can still be constrained with a cloud deck at 1~mbar, while H$_2$O and CH$_4$ constraints are still possible with a 0.1~mbar cloud deck. 
    
\end{itemize}

We note that throughout this work, we have assumed a worst-case scenario of a grey cloud opacity. In reality, this may not be the case as clouds may have a non-gray opacity with lower impact on the spectrum at longer wavelengths. Our findings are therefore conservative in that regard, as any reduction in cloud opacity will lead to improved constraints on atmospheric parameters, provided such considerations are included in retrieval models. In principle, the shorter wavelengths below $\sim$1~$\mu$m accessible with NIRISS could provide additional constraints on the scattering slope due to clouds/hazes. Additionally, the NIRSpec G395H spectral range contains some opacity windows between $\sim$4-5 $\mu$m   in the absence of other molecular absorption (e.g. CO or CO$_2$) beyond those considered here. Such opacity windows could also provide constraints on contributions of clouds/hazes to the spectral continuum in the infrared.

We have also assumed that the planet's terminator is fully covered by clouds, again as a worst-case scenario. In reality there may only be partial cloud coverage, resulting in the troughs between spectral features lying on a continuum from ``V''-shaped in the cloud-free case to ``U''-shaped in the fully cloudy case. With high-precision JWST spectra, atmospheric retrievals would enable accurate inferences of both the chemical abundances as well as the cloud parameters.

All our retrievals have been carried out on simulated observations at their native resolutions. We note however that it is normal practice to bin spectra down to a specific resolution. As we mention in section \ref{sec:benchmark_canonical}, excessive binning may result in loss of information, particularly if there are very high-altitude clouds present, leading to poorly constrained or unconstrained abundance estimates. However, we find that binned NIRSpec spectra with R$\gtrsim$500 provide constraints approaching those obtained with native resolution spectra for our limiting case with a 0.1 mbar cloud deck in TOI-732~c. While our focus throughout this work has been on the high-resolution NIRSpec gratings, NIRSpec also offers medium-resolution equivalents. Given our findings, the medium-resolution gratings are expected to yield constraints that are comparable to those from the high-resolution gratings particularly for atmospheres that are cloud-free or with low-altitude cloud decks.

K2-18~b in particular is set to be extensively observed during Cycle 1 as part of GO programs 2372 and 2722 (PI: Renyu Hu and Nikku Madhusudhan, respectively). Program 2372 is set to observe 2 transits with NIRSpec G235H and 4 with NIRSpec G395H, while program 2722 will combine observations from NIRISS, NIRSpec G395H and MIRI LRS, allocating one transit to each. Given both programs will achieve an extensive wavelength coverage, both can be expected to lead to abundance constraints for H$_2$O, CH$_4$ and NH$_3$ with precisions comparable to or better than those we find in this work. Moreover, CO and CO$_2$, which may be present in trace amounts, could also be detectable under certain conditions. This is thanks to the large number of transits set to be observed with NIRSpec G395H, which offers a wavelength range encompassing the prominent spectral features of both molecules.

It can be noted that throughout our results, the mixing ratios of CH$_4$ and NH$_3$ are more readily constrained than that of H$_2$O. This is evident from both the tighter abundance constraints obtained for CH$_4$ and NH$_3$ compared to H$_2$O with the canonical model in section \ref{sec:wavelength_coverage}, as well in section \ref{sec:benchmark_canonical}, where the retrieved constraints for CH$_4$ and NH$_3$ deteriorate less rapidly than those for H$_2$O with increasing cloud deck altitudes. Even when CH$_4$ and NH$_3$ were depleted by 1~dex in section \ref{sec:benchmark_depleted}, CH$_4$ remained similarly detectable compared to the significantly more abundant H$_2$O. This is in contrast to the failure so far in detecting either CH$_4$ or NH$_3$ in a temperate exoplanet atmosphere, the so-called ``Missing Methane'' Problem \citep{Stevenson2010, Madhu2011, Madhusudhan2020}. Given the ease with which both CH$_4$ and NH$_3$ can be detected, it is evident that JWST is crucial in resolving this problem.

In this work, we focus on the two mini-Neptunes K2-18~b and TOI-732~c, in both cases using atmospheric compositions where H$_2$O is always at super-solar abundances, while CH$_4$ and NH$_3$ are at either solar or super-solar abundances. In reality, compositions that significantly deviate from those considered here, such as a more extreme depletion of CH$_4$ and NH$_3$, are possible and would be consistent with current HST observations of K2-18~b \citep{Benneke2019, Madhusudhan2020}. Such depletions may either lead to even more precise constraints, by further alleviating degeneracies arising from overlapping spectral features as seen in section \ref{sec:benchmark_depleted}, or in more extreme cases, make CH$_4$ and NH$_3$ harder to detect as their spectral contributions may be masked by stronger H$_2$O features.

Our study benchmarks and showcases the ability of JWST to constrain atmospheric abundances and other properties of temperate mini-Neptunes with unprecedented precision, vastly outperforming what is currently possible with HST. This is thanks to a generational improvement in wavelength coverage, sensitivity and resolution, which together allow for atmospheres to be characterised even with high-altitude clouds and in the absence of supporting data in the optical.

\section*{Acknowledgements}
 We thank the anonymous referee for their valuable review and feedback. We additionally thank Luis Welbanks and Subhajit Sarkar for helpful discussions. This work was performed using resources provided by the Cambridge Service for Data Driven Discovery (CSD3) operated by the University of Cambridge Research Computing Service (\url{www.csd3.cam.ac.uk}), provided by Dell EMC and Intel using Tier-2 funding from the Engineering and Physical Sciences Research Council (capital grant EP/P020259/1), and DiRAC funding from the Science and Technology Facilities Council (\url{www.dirac.ac.uk}).

\section*{Data Availability}

No new data were generated or analysed in support of this research.


\bibliographystyle{mnras}
\bibliography{ms}

\bsp	
\label{lastpage}
\end{document}